\documentclass[journal,10pt,twoside]{IEEEtranTCOM}
\normalsize
\pdfoutput=1
\newif\ifonecolumn
\usepackage[final]{graphicx}
\usepackage{amsfonts}
\usepackage{epsfig}
\usepackage{amssymb}
\usepackage{amsmath}

\usepackage{psfrag}
\usepackage{balance}
\usepackage{tikz}
\usetikzlibrary{calc}
\usetikzlibrary{positioning}
\usepackage{pgfplots}
\usepackage{verbatim}
\usepackage{mathtools}

\newcommand{\beq}{\begin{equation}}
\newcommand{\eeq}{\end{equation}}
\newcommand{\ben}{\begin{enumerate}}
\newcommand{\een}{\end{enumerate}}
\newcommand{\bit}{\begin{itemize}}
\newcommand{\eit}{\end{itemize}}

\newcommand{\abs}[1]{\left\lvert#1\right\rvert}
\newcommand{\set}[1]{\left\{#1\right\}}
\newcommand{\norm}[1]{\left\lVert#1\right\rVert}
\newcommand{\rx}{\mathrm{x}}
\newcommand{\ra}{\mathrm{a}}
\newcommand{\rb}{\mathrm{b}}
\newcommand{\rc}{\mathrm{c}}

\renewcommand{\a}{\mathbf{a}}
\renewcommand{\c}{\mathbf{c}}
\newcommand{\bom}{{\boldsymbol{\omega}}}
\newcommand{\Ca}{C_\mathrm{a}}
\newcommand{\Cb}{C_\mathrm{b}}
\newcommand{\Cc}{C_\mathrm{c}}
\newcommand{\Cx}{C_\mathrm{x}}
\newcommand{\xa}{\mathbf{x}_\mathrm{a}}
\newcommand{\xb}{\mathbf{x}_\mathrm{b}}
\newcommand{\xc}{\mathbf{x}_\mathrm{c}}

\newcommand{\x}{\mathbf{x}}

\newcommand{\xach}{\mathbf{x}_{\mathrm a}^{\mathrm{ch}}}
\newcommand{\xbch}{\mathbf{x}_{\mathrm b}^{\mathrm{ch}}}

\newcommand{\xap}{\mathbf{x}_{\mathrm a}^{\mathrm{p}}}
\newcommand{\xbp}{\mathbf{x}_{\mathrm b}^{\mathrm{p}}}
\newcommand{\xxp}{\mathbf{x}_{\mathrm x}^{\mathrm{p}}}
\newcommand{\xm}{\mathbf{x}^{(m)}}
\newcommand{\y}{\mathbf y}
\newcommand{\Ex}{E_{\mathrm x}}
\newcommand{\f}{\mathbf{f}}
\newcommand{\dotFPi}{\conic(\dot{\mathcal Q}_{\Pi, \Pi_{\rm c}})}
\newcommand{\dotF}{\dot{\mathcal F}_{\Pi, \Pi_{\rm c}}}
\newcommand{\QPi}{\mathcal Q_{\Pi, \Pi_{\rm c}}}
\newcommand{\QfPi}{\mathcal Q^{\mathbf{f}}_{\Pi, \Pi_{\rm c}}}
\newcommand{\dotQPi}{\dot{\mathcal Q}_{\Pi, \Pi_{\rm c}}}
\newcommand{\tQPi}{\tilde{\mathcal Q}_{\Pi,\Pi_{\rm c}}}
\newcommand{\dmin}{d_\mathrm{min}}

\newcommand{\hmin}{h_\mathrm{min}}
\newcommand{\wbar}{\bar{w}}
\newcommand{\Fsec}{{\mathcal F}_{\rm sec}}

\newtheorem{example}{Example}
\newtheorem{lemma}{Lemma}

\newtheorem{definition}{Definition}

\newtheorem{proposition}{Proposition}

\DeclareMathOperator{\conv}{conv}
\DeclareMathOperator{\conic}{conic}
\DeclareMathOperator{\proj}{proj}
\DeclareMathOperator{\argmax}{arg\,max}

\newcommand\T{\rule{0pt}{2.6ex}}
\newcommand\B{\rule[-1.2ex]{0pt}{0pt}}

\begin{document}
\title{Minimum Pseudoweight Analysis of 3-Dimensional Turbo Codes}%
\author{Eirik~Rosnes,~\IEEEmembership{Senior~Member,~IEEE}, Michael~Helmling, and Alexandre~Graell~i~Amat,~\IEEEmembership{Senior~Member,~IEEE}%
%
\thanks{The work of E.\ Rosnes and M.\ Helmling was partially funded by the DFG (grant RU-1524/2-1) and by DAAD / RCN (grant 54565400 within the German-Norwegian Collaborative Research Support Scheme). The work of A.\ Graell~i~Amat was supported by the Swedish Research Council under grant \#2011-596. The material in this paper was presented in part at the 2011 IEEE International Symposium on Information Theory, St.\ Petersburg, Russia, Jul./Aug.\ 2011.}%
\thanks{E.\ Rosnes was with Ceragon Networks AS, Kokstadveien 23, N-5257 Kokstad, Norway. He is now with the Selmer Center, Department of Informatics, University of Bergen, N-5020 Bergen, Norway, and the Simula Research Lab. E-mail: eirik@ii.uib.no.}%
\thanks{M.\ Helmling was with the Department of Mathematics, University of  Kaiserslautern, 67663 Kaiserslautern, Germany. He is now with the Mathematical Institute, University of Koblenz-Landau, 56070 Koblenz, Germany. Email: helmling@uni-koblenz.de.}%
\thanks{A.\ Graell i Amat is with the Department of Signals and Systems, Chalmers University of Technology, SE-412 96 Gothenburg, Sweden. Email: alexandre.graell@chalmers.se.}}%
%

\maketitle

\begin{abstract}
In this work, we consider pseudocodewords of (relaxed) linear programming (LP) decoding  of $3$-dimensional turbo codes (3D-TCs). 
We present a relaxed LP decoder for 3D-TCs, adapting the relaxed LP decoder for conventional turbo codes proposed by Feldman in his thesis. 
We show that the 3D-TC polytope is \emph{proper} and \emph{$C$-symmetric}, and make a connection to finite graph covers of the 3D-TC factor graph. This connection is used to show that the support set of any pseudocodeword is a stopping set of iterative decoding of 3D-TCs using maximum \emph{a posteriori} constituent decoders on the binary erasure channel. Furthermore, we compute ensemble-average pseudoweight enumerators of 3D-TCs and perform a finite-length minimum pseudoweight analysis for small cover degrees. Also, an explicit description of the fundamental cone of the 3D-TC polytope is given. 
 Finally, we present an extensive numerical study of small-to-medium block length 3D-TCs, which shows that 1) typically (i.e., in most cases) when the minimum distance $d_{\rm min}$ and/or the stopping distance $h_{\rm min}$ is high, the minimum pseudoweight (on the additive white Gaussian noise channel) is strictly smaller than both the $d_{\rm min}$ and the $h_{\rm min}$, and 2) the minimum pseudoweight grows with the block length, at least for small-to-medium block lengths.
%
\end{abstract}


\begin{keywords}
3-dimensional (3D) turbo codes, graph cover, hybrid concatenated codes, linear programming (LP) decoding, QPP interleavers, pseudocodewords, pseudoweight, turbo codes, uniform interleaver.
\end{keywords}


\section{Introduction} \label{sec:intro}

\IEEEPARstart{T}{urbo} codes (TCs) have gained considerable attention since their
introduction by Berrou \emph{et al.} in 1993 \cite{ber93} due to
their near-capacity performance and low decoding complexity. The
conventional TC is a parallel concatenation of two identical
recursive systematic convolutional encoders separated by a
pseudorandom interleaver. To improve the performance of TCs in the error floor region,
hybrid concatenated codes (HCCs) can be used. In
\cite{ber09}, a powerful HCC nicknamed
\mbox{3-dimensional} turbo code (3D-TC) was introduced. The coding scheme consists of
a conventional turbo encoder and a \emph{patch}, where  a
fraction $\lambda$ of the parity bits from the turbo encoder are 
post-encoded by a third rate-$1$ encoder. The value of $\lambda$ can
be used to trade off performance in the waterfall region with
performance in the error floor region. As shown in \cite{ber09},
this coding scheme is able to provide very low error rates for a
wide range of block lengths and code rates at the expense of a small
increase in decoding complexity with respect to conventional TCs. In a recent work \cite{GraRos11}, an in-depth performance analysis of 3D-TCs was conducted. Stopping sets for 3D-TCs were treated in \cite{GraRosIsit2010}. Finally, we remark that tuned TCs \cite{kol12} are another family of HCC 
ensembles where performance in the waterfall and error floor regions can be traded off using a tuning parameter.

The use of linear programming (LP) to decode turbo-like codes was introduced by Feldman \emph{et al.} \cite{Feldman+02LPTurboDecoding,fel03}. They proposed an LP formulation that resembles a shortest path problem, based on the trellis graph representation of the constituent convolutional codes. The natural polytope for LP decoding would be the convex hull of all codewords, in which case LP decoding is equivalent to maximum-likelihood (ML) decoding. However, an efficient description of that polytope is not known in general, i.e., its description length most likely grows exponentially with the block length. The  formulation proposed by Feldman \emph{et al.} \cite{Feldman+02LPTurboDecoding,fel03}, which grows only linearly with the block length, is a relaxation in the sense that it describes a superset of the ML decoding polytope, introducing additional, fractional vertices. The vertices of the relaxed polytope (both integral and fractional) are what the authors called \emph{pseudocodewords} in \cite{fel05}.

The same authors also introduced a different LP formulation to decode low-density parity-check (LDPC) codes \cite{fel03,fel05} that has been extensively studied since then by various researchers. For that LP decoder, Vontobel and Koetter \cite{koe05} showed that the set of the polytope's points with rational coordinates (which in particular includes all of its vertices) is equal to the set of all pseudocodewords derived from all finite graph covers of the Tanner graph. In \cite{ros06}, a similar result was established (but with no proof included) for the case of conventional TCs. Here, we prove that statement for the case of 3D-TCs. 

In this work, we study (relaxed) LP decoding of 3D-TCs, explore the connection to finite graph covers of the 3D-TC factor graph \cite{ksc01}, and adapt the concept of a pseudocodeword. Furthermore, we compute ensemble-average pseudoweight enumerators and perform a finite-length minimum  additive white Gaussian noise (AWGN) pseudoweight analysis which shows that the minimum AWGN pseudoweight grows with the block length, at least for small-to-medium block lengths. 
%
%
Furthermore, we show by several examples and by computer search that typically (i.e., in most cases) when  the minimum distance  $d_{\rm min}$ and/or the stopping distance $h_{\rm min}$ is high, the minimum AWGN pseudoweight, denoted by $w^{\rm AWGN}_{\rm min}$, is strictly smaller than both the $\dmin$ and the $\hmin$ for these codes. In \cite{che11}, Chertkov and Stepanov presented an updated, more efficient (compared to the algorithm from \cite{che08}) minimum pseudoweight search algorithm based entirely on the concept of the \emph{fundamental cone} \cite{koe05} of the LDPC code LP decoder. We will show that such a fundamental cone can be described also for the 3D-TC LP decoder.

Some other work related to the content of this paper is the work on pseudocodeword analysis of binary and nonbinary (protograph-based) LDPC and generalized LDPC codes. See, for instance, \cite{abu11,div12,div11}, and references therein. For such codes, the component codes are the trivial repetition code and single parity-check codes, or in the case of generalized LDPC codes, more advanced classical linear block codes. However, not much work has been done on pseudocodeword analysis for turbo-like codes with trellis-based constituent codes. In contrast to these previous works, the trellis-based approach in this paper is different and provides a pseudocodeword analysis of 3D-TCs that can be adapted also to other trellis-based turbo-like codes or  concatenated codes based on two or more convolutional codes. We remark that some results on enumerating pseudocodewords for convolutional codes have already been provided by Boston and Conti in \cite{bos09,con11}. Finally, it is worth mentioning \cite{von13} which presents, among several results, a combinatorial characterization of the \emph{Bethe entropy function} in terms of finite graph covers of the factor graph under consideration. In particular, a characterization in terms of the average number of preimages of a \emph{pseudomarginal} vector of rational entries. 
In contrast to the general framework in \cite{von13}, 
this paper discusses techniques to numerically deal with large-degree function nodes representing the indicator function of (long) convolutional codes.

The remainder of the paper is organized as follows. In Section~\ref{sec:coding scheme}, we describe the system model and introduce some notation. 
In Section~\ref{sec:lp_decoding}, we describe (relaxed) LP decoding of 3D-TCs. The connection to finite graph covers of the 3D-TC factor graph is explored in Section~\ref{sec:cover}. Ensemble-average pseudoweight enumerators of 3D-TCs are computed in Section~\ref{sec:average_analysis}.  These enumerators are subsequently used to perform a probabilistic  finite-length minimum AWGN pseudoweight analysis of 3D-TCs. In Section~\ref{sec:searching}, an efficient heuristic for searching for low AWGN pseudoweight pseudocodewords is discussed. Finally, in Section~\ref{sec:numerical}, an extensive numerical study is presented, and some conclusions are drawn in Section~\ref{sec:conclu}.

\section{Coding Scheme} \label{sec:coding scheme}
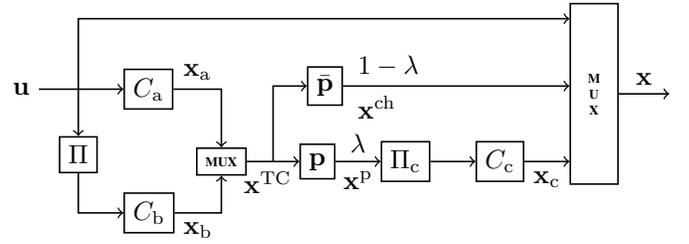
\begin{figure}[t!]
\par
\begin{center}
%
\begin{tikzpicture}[sq/.style={rectangle,draw,semithick},arrow/.style={->,semithick},node distance=5mm and 6mm]
        \node (u) at (0,0) {$\mathbf u$};
        \node (cross) [right=of u,xshift=-2mm] {};
        \node[sq] (pi) [below=of cross] {$\Pi$};
        \node[sq] (Ca) [right=of cross.center] {$C_\mathrm a$};
        \node[sq] (Cb) [below right=of pi.center] {$C_\mathrm b$};
        \node[sq] (mux1) [below right=of Ca,xshift=-3mm,font=\tiny\bfseries] {MUX};
        \node[sq] (p) [right=of mux1,xshift=1mm] {$\mathbf p$};
        \node[sq] (pbar) [above=of p,xshift=1mm] {$\bar{\mathbf p}$};
        \node[sq] (pic) [right=of p] {$\Pi_\mathrm c$};
        \node[sq] (Cc) [right=of pic] {$C_{\mathrm c}$};
        \node[sq] (mux2) [right=of Cc,minimum height=2.4cm,yshift=0.9cm,text centered,text width=4mm,font=\tiny\bfseries] {M\\U\\X};
        \draw[arrow] (u) -- (Ca);
        \draw[arrow,<->] (pi) |- ($ (mux2.west) + (0,1) $);
        \draw[arrow] (p) -- node[below] {$\mathbf x^\mathrm p$} node[above] {$\lambda$} (pic);
        \draw[arrow] (pic) -- (Cc);
        \draw[arrow] (Ca) -| node[near start,above] {$\mathbf x_\mathrm a$} (mux1);
        \draw[arrow] (pi) |- (Cb);
        \draw[arrow] (Cb) -| node[near start,below] {$\mathbf x_\mathrm b$} (mux1);
        \draw[arrow] (mux1) -- node[below,pos=.4] {$\mathbf x^{\mathrm{TC}}$} (p);
        \draw[arrow] ($ (mux1.east)!.5!(p.west) $) |- (pbar);
        \draw[arrow] (pbar.east) -- node[below,pos=.15] {$\mathbf x^{\mathrm{ch}}$} node[pos=.2,above] {$1-\lambda$} (pbar.east -| mux2.west);
        \draw[arrow] (Cc) -- node[below] {$\mathbf x_{\mathrm c}$} (Cc.east -| mux2.west);
        \draw[arrow] (mux2) -- node[auto]{$\mathbf x$} +(1,0);
    \end{tikzpicture}
\end{center}
\caption{\label{fig:encoder} {3D turbo encoder. A fraction
$\lambda$ of the parity bits from both constituent encoders $\Ca$
and $\Cb$ are grouped by a parallel/serial multiplexer, permuted by the interleaver
$\Pi_{\rm c}$, and encoded by the rate-$1$ post-encoder $\Cc$.}}
\end{figure}

A block diagram of the 3D-TC is depicted in
Fig.~\ref{fig:encoder}. The information data sequence $\mathbf{u}$
of length $K$ bits is encoded by a binary conventional turbo
encoder. By a conventional turbo encoder we mean the parallel
concatenation of two identical rate-$1$ recursive convolutional
encoders, denoted by $\Ca$ and $\Cb$, respectively. Here, $\Ca$ and
$\Cb$ are 8-state recursive convolutional encoders with generator
polynomial $g(D)=(1+D+D^3)/(1+D^2+D^3)$, i.e., the 8-state
constituent encoder specified in the 3GPP LTE standard
\cite{gpplte}. The code sequences of $\Ca$ and $\Cb$ are denoted by
$\xa$ and $\xb$, respectively. We also denote by $\x^{\mathrm{TC}}$
the codeword obtained by alternating bits from $\xa$ and $\xb$. A
fraction $\lambda$ ($0\leq \lambda \leq 1$), called the
\textit{permeability rate}, of the parity bits from
$\x^{\mathrm{TC}}$ are permuted by the interleaver $\Pi_{\rm c}$ (of
length $N_\mathrm{c}=2\lambda K$), and encoded by an encoder of
unity rate $\Cc$  with generator polynomial $g(D)=1/(1+D^2)$, called the \textit{patch} or the
\textit{post-encoder} \cite{ber09}. This can be properly represented
by a puncturing pattern $\mathbf{p}$ applied to $\x^{\mathrm{TC}}$
(see Fig.~\ref{fig:encoder}) of period $N_{\rm p}$ containing
$\lambda N_{\rm p}$ ones (where a one means that the bit is not
punctured).  Note that the encoder of the patch is like two
  accumulators, one operating on the even bits and one operating on the odd
  bits. 
The fraction $1-\lambda$ of parity bits which are not
encoded by $\Cc$ is sent directly to the channel. Equivalently, this
can be represented by a puncturing pattern $\bar{\mathbf{p}}$, the
complement of $\mathbf{p}$. We denote by $\xc$ the code sequence produced by
$\Cc$. Also, we denote by $\xa^\mathrm{ch}$ and $\xb^{\mathrm{ch}}$
the \textit{sub-codewords} of $\xa$ and $\xb$, respectively, sent
directly to the channel, and by $\x^\mathrm{ch}$ the codeword
obtained by multiplexing (in some order) the bits from $\xa^\mathrm{ch}$ and
$\xb^\mathrm{ch}$. Likewise, we denote by $\xa^{\mathrm{p}}$ and
$\xb^{\mathrm{p}}$ the \textit{sub-codewords} of $\xa$ and $\xb$,
respectively, encoded by $\Cc$, and by $\x^{\mathrm{p}}$ the
codeword obtained by multiplexing (in some order) the bits from $\xa^\mathrm{p}$ and
$\xb^\mathrm{p}$. Finally, the information sequence and the code
sequences $\x^{\mathrm{ch}}$ and $\xc$ are multiplexed to form the
code sequence $\mathbf{x}$, of length $N$ bits, transmitted to the
channel. Note that the overall nominal code rate of the 3D-TC is
$R=K/N=1/3$, the same as for the conventional TC without the
patch. Higher code rates can be obtained either by puncturing
$\x^{\mathrm{ch}}$ or by puncturing the output of the patch, $\xc$.

In \cite{ber09}, regular puncturing patterns of period $2/\lambda$
were considered for $\mathbf{p}$. For instance, if $\lambda=1/4$,
every fourth bit from each of the encoders of the outer TC
are encoded by encoder $\Cc$. The remaining bits are sent directly
to the channel, and it follows that $\mathbf{p}=[11000000]$ and
$\bar{\mathbf{p}}=[00111111]$. 
Note that with this particular puncturing pattern and even with the generator polynomial $g(D) = 1 / (1+D^2)$ for $C_{\rm c}$ (which is like two separate accumulators operating on even and odd bits, respectively), the  bit streams $\x_{\rm a}$ and $\x_{\rm b}$ are in general intermingled because of the  interleaver $\Pi_{\rm c}$.

\section{LP Decoding of 3D-TCs} \label{sec:lp_decoding}

In this section, we consider relaxed LP decoding of 3D-TCs, adapting the relaxation proposed in \cite{fel03} for conventional TCs to 3D-TCs.  

Let $T_\rx = T_{\rm x}(V_{\rm x},E_{\rm x})$ denote the  \emph{information bit-oriented trellis} of  $C_{\rm x}$, ${\rm x}={\rm a},{\rm b},{\rm c}$, where the vertex set $V_{\rm x}$ partitions 
as $V_{\rm x} = \cup_{t=0}^{I_{\rm x}} V_{{\rm x},t}$, which also induces the partition  $E_{\rm x} = \cup_{t=0}^{I_{\rm x}-1} E_{{\rm x},t}$ of the edge set $E_{\rm x}$, where $I_{\rm x}$ is the trellis length of $T_{\rm x}$. In the following, the encoders $C_{\rm x}$ (and their corresponding trellises $T_{\rm x}$) are assumed (with some abuse of notation) to be systematic, in the sense that the output bits are prefixed with the input bits. Thus, $C_{\rm  x}$ is regarded as a rate-$1/2$ encoder, and the trellis $T_{\rm x}$ has an output label containing two bits, for ${\rm x}={\rm a},{\rm b},{\rm c}$. Now, let $e \in E_{{\rm x},t}$ be an arbitrary edge from the $t$th trellis section. The $i$th bit in the output label of $e$ is denoted by  $c_i(e)$, $i=0,\dots,n_{{\rm x},t}-1$, the starting state of $e$ as $s^S(e)$, and 
the ending state of $e$ as $s^E(e)$, where $n_{{\rm x},t}$ is the number of bits in the output label of an edge $e \in E_{{\rm x},t}$.

For $\rm x=a,b,c$, we define the path polytope $\mathcal Q_\rx$ to be the set of all $\f^\rx \in [0,1]^{\abs{E_\rx}}$ satisfying
\begin{IEEEeqnarray}{rCl?l}
  \sum_{e \in E_{\rx,0}} f^\rx_e
    &=& 1
    \label{eq:flowSource}
  \\
  \sum_{\substack{e \in E_\rx:\\s^S(e)=v}} f^\rx_e
    &=& \sum_{\substack{e \in E_\rx:\\ s^E(e)=v}} f^\rx_e
      & \text{for all } v\in V_{\rx,t}\label{eq:flowConservation}
  \\[-7mm]
    &&& \text{and } t = 1,\dotsc, I_\rx-1 \IEEEnonumber
\end{IEEEeqnarray}
and let $\mathcal Q = \mathcal Q_\ra \times \mathcal Q_\rb \times \mathcal Q_\rc$. Note that $\mathcal Q$ is the set of all feasible network flows through the three trellis graphs $T_\rx$, $\rx = \ra,\rb,\rc$.

Next, we define the polytope $\QPi$ as the pairs $(\tilde\y, \f)$, where $\tilde\y \in [0,1]^{N+2\lambda K}$ and $\f=(\f^\ra,\f^\rb,\f^\rc) \in [0,1]^{\abs{E_\ra \cup E_\rb \cup E_\rc}}$, meeting the constraints
\begin{IEEEeqnarray*}{rCl?l}
  (\f^\ra, \f^\rb, \f^\rc)
    &\in& \mathcal Q
  \\
  \sum_{\substack{e \in E_{\rx, t}:\\c_i(e)=1}} f_e^\rx
    & = & \tilde y_{\rho_\rx(\phi_\rx(t,i))}
        & \text{for }t=0, \dotsc, I_\rx-1,\IEEEyesnumber\label{eq:flowY}
  \\[-7mm]
    & & & i=0, \dotsc, n_{\rx,t}-1,\text{ and}
  \\
    & & & \rx=\ra,\rb,\rc.    
\end{IEEEeqnarray*}
Here, $\phi_\rx(t,i)=\sum_{j=0}^{t-1} n_{\rx,j}+i$, and $\rho_\rx(\cdot)$\label{page:rhox} denotes the mapping of codeword indices of the constituent encoder $\Cx$ to codeword indices of the overall codeword of the 3D-TC appended with the $2\lambda K$ parity bits from encoders $\Ca$ and $\Cb$ which are sent to the patch.

Finally, let
\begin{IEEEeqnarray*}{rCl}
  \dotQPi
    &=& \Big\{\y \in [0,1]^N:\:\exists \hat\y \in [0,1]^{2\lambda K},
        \f \in \mathcal Q
  \\
    & & \quad\:\text{with }((\y, \hat\y), \f) \in \QPi\Big\}
\end{IEEEeqnarray*}
be the projection of $\QPi$ onto the first $N$ variables.

Relaxed LP decoding (on a binary-input memoryless channel)  of 3D-TCs  can be described by the linear program
\begin{equation}
  \text{minimize } \sum_{l=0}^{N-1} \lambda_l y_l
  \text{ subject to } \y \in \dotQPi
  \label{eq:LP1}
\end{equation}
%
where
\begin{displaymath}
\lambda_l = \log \left( \frac{{\rm Pr}\{r_l | c_l=0\}}{{\rm Pr}\{r_l | c_l=1 \}} \right),\; l=0,\dots,N-1,  
\end{displaymath}
$c_l$ is the $l$th codeword bit, and $r_l$ is the $l$th component of the received vector. If instead of $\dotQPi$ we use the convex hull of the codewords of the 3D-TC, then solving the linear program in (\ref{eq:LP1}) is equivalent to ML decoding.


The notion of a \emph{proper} and \emph{$C$-symmetric} polytope was introduced in \cite[Ch.\ 4]{fel03} where the author proved that the probability of error of LP decoding  is independent of the transmitted codeword on a binary-input output-symmetric memoryless channel when the underlying code is linear and the polytope is proper and $C$-symmetric.

\begin{proposition} \label{lem:1}
Let $C$ denote a given 3D-TC with interleavers
$\Pi$ and $\Pi_\rc$. 
The polytope $\dotQPi$  is proper, i.e., $\dotQPi \cap \{0,1\}^N = C$ and $C$-symmetric,
 i.e., for any $\y \in \dotQPi$ and $\mathbf{c} \in C$ it holds that $\abs{\y-\mathbf{c}} \in  \dotQPi$.
\end{proposition}

Feldman proved a similar statement in the context of LDPC codes (Lemma~5.2 and Theorem~5.4 in \cite{fel03}). However, note that his proof is based explicitly on the inequalities of the LP formulation for LDPC codes, and therefore does not generalize to the polytope $\dotQPi$.

In Appendix~\ref{sec:AppendixA}, we give a formal proof of both statements in a much more general setting.

\section{Finite Graph Covers} \label{sec:cover}
Let $C$ denote a given 3D-TC with interleavers $\Pi$ and $\Pi_{\rm c}$,   and constituent codes  $C_{\rm x}$, ${\rm x}={\rm a},{\rm b},{\rm c}$. The factor graph \cite{ksc01} of $C_{\rm x}$, denoted by $\Gamma(C_{\rm x})$, is composed of \emph{state}, \emph{input}, \emph{parity}, and \emph{check} vertices. The state vertices $s_{{\rm x},0},\dots,s_{{\rm x},I_{\rm x}}$ in $\Gamma(C_{\rm x})$ represent state spaces of the length-$I_{\rm x}$ information bit-oriented trellis $T_{\rm x}$ of $C_{\rm x}$. The $l$th check vertex represents the $l$th trellis section, i.e., it  is an indicator function for the set of allowed combinations of \emph{left} state, input symbol, parity symbol, and \emph{right} state.
A factor graph $\Gamma(C)$ of $C$ is constructed as follows.
\begin{enumerate}
\item Remove all the input vertices of $\Gamma(C_{\rm b})$ by connecting the $l$th input vertex of $\Gamma(C_{\rm a})$ to the $\Pi(l)$th check vertex of $\Gamma(C_{\rm b})$.  
\item Remove all the input vertices of $\Gamma(C_{\rm c})$ by connecting the parity vertex (from either $\Gamma(C_{\rm a})$ or $\Gamma(C_{\rm b})$) corresponding to the $l$th bit in $\mathbf{x}^{\rm p}$   
to the $\Pi_{\rm c}(l)$th check vertex of $\Gamma(C_{\rm c})$.
\end{enumerate}

\begin{figure}[htb]
\par
\begin{center}
\includegraphics[width=\columnwidth]{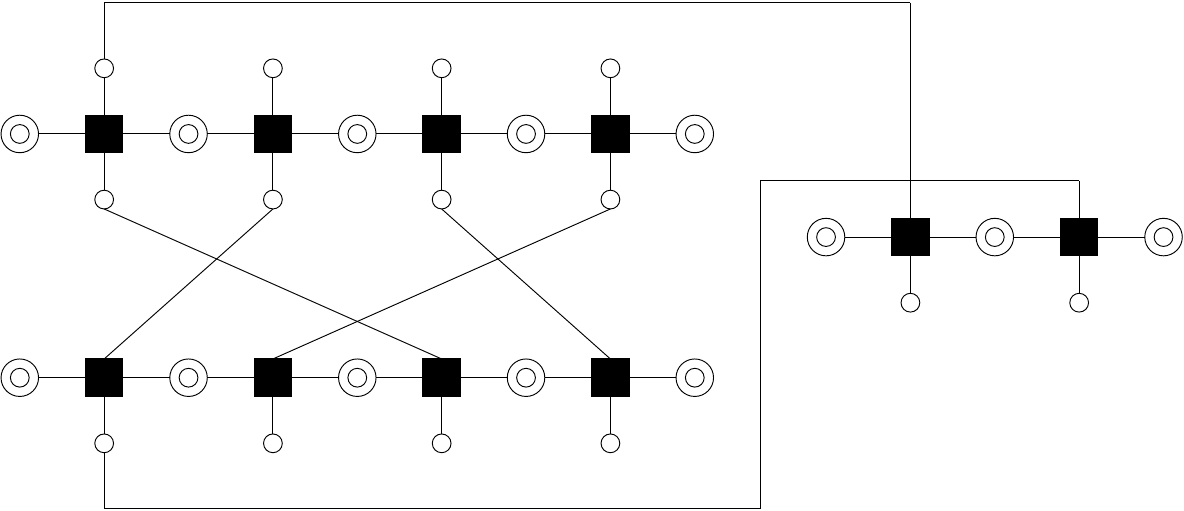}
\end{center}
\caption{\label{cover3} {Factor graph of a nominal rate-$1/3$ 3D-TC with $\lambda=1/4$, using the regular puncturing pattern $\mathbf{p}=[11000000]$.}}
 \end{figure}

\begin{figure*}[htb]
\par
\begin{center}
\includegraphics[width=2\columnwidth]{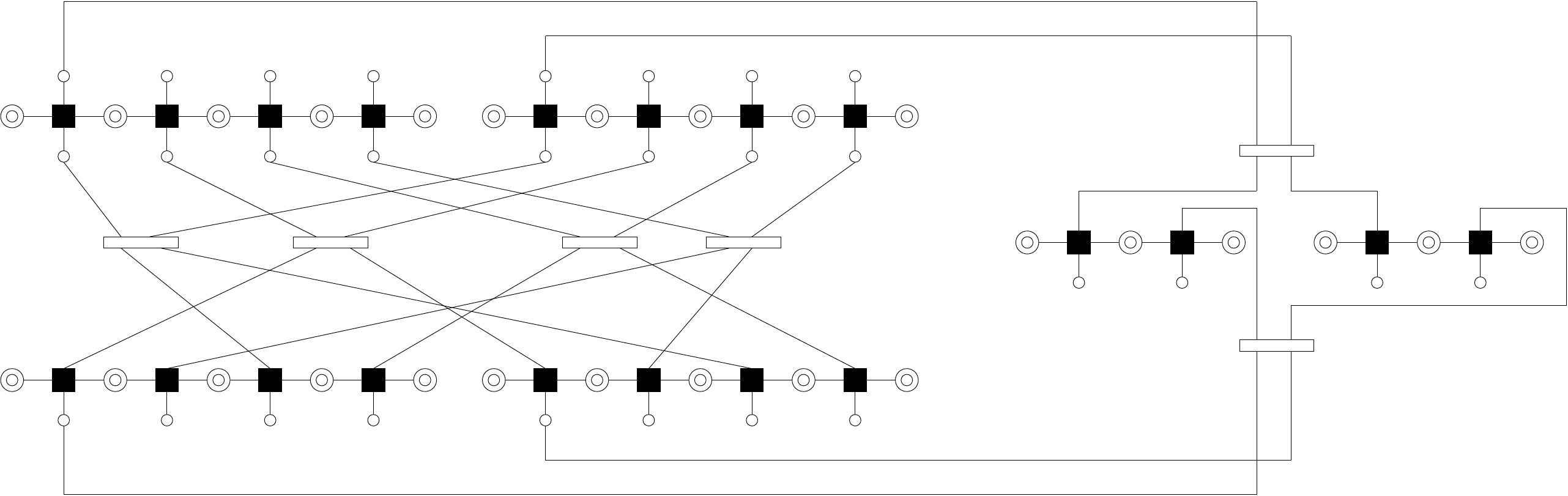}
\end{center}
\caption{\label{cover4} {Degree-$2$ cover of the factor graph in Fig.~\ref{cover3}. The six small rectangular boxes are permutations of size $2$ that can be chosen arbitrarily.}}
\end{figure*}

To construct a degree-$m$ cover of $\Gamma(C)$, denoted by $\Gamma^{(m)}(C)$, we first make  $m$ identical copies of $\Gamma(C)$.  
%
Now, any permutation of the edges,  denoted by $E=E(\Gamma^{(m)}(C))$, connecting the copies of the constituent factor graphs $\Gamma(C_{\rm x})$ such that the following conditions are satisfied, will give a valid cover of  $\Gamma(C)$. 
\begin{enumerate}
\item The $m$ copies of the $l$th input vertex of $\Gamma(C_{\rm a})$  should be connected by a one-to-one mapping (or permutation) to the $m$ copies of the $\Pi(l)$th check vertex of $\Gamma(C_{\rm b})$.
\item The $m$ copies  of the parity vertex  (from either $\Gamma(C_{\rm a})$ or $\Gamma(C_{\rm b})$) corresponding to the $l$th bit in $\mathbf{x}^{\rm p}$ should be connected  by a permutation  to the $m$ copies of the $\Pi_{\rm c}(l)$th check vertex of $\Gamma(C_{\rm c})$.
%
%
\end{enumerate}

 The corresponding code is denoted by $C^{(m)}$. 
%
%
Let $\mathbf{x}^{(m)}=(x_{0}^{(0)},\dots,x_{N-1}^{(0)},\dots,x_{0}^{(m-1)},\dots,x_{N-1}^{(m-1)})$ denote a codeword in $C^{(m)}$, 
define 
\begin{displaymath}
\omega_l(\mathbf{x}^{(m)})=\frac{|\{i:\:x_{l}^{(i)}=1\}|}{m}
\end{displaymath}
and let $\boldsymbol{\omega}=\boldsymbol{\omega}(\mathbf{x}^{(m)})=(\omega_0(\mathbf{x}^{(m)}),\dots,\omega_{N-1}(\mathbf{x}^{(m)}))$. Now, $\boldsymbol{\omega}$ as defined above is said to  be a \emph{graph-cover pseudocodeword} of degree $m$. 

\begin{example} \label{ex:1}
 Fig.~\ref{cover3} depicts the factor graph of a nominal rate-$1/3$ 3D-TC with $\lambda=1/4$ and input length $K=4$, using the regular puncturing pattern $\mathbf{p}=[11000000]$.  The upper part to the left  is the factor graph of  $C_{\rm a}$, the lower part to the left is the factor graph  of $C_{\rm b}$, and the right part is the factor graph of $C_{\rm c}$. The black squares in the graph are check vertices corresponding to trellis sections. The single circles are input and parity vertices, and the double circles are state vertices.  Fig.~\ref{cover4} depicts a degree-$2$ cover of the factor graph from Fig.~\ref{cover3}. The six small rectangular boxes are permutations of size $2$ that can be chosen arbitrarily.
\end{example}

\begin{proposition} \label{lem:2}
The following statements are true:
\begin{enumerate}
\item The points in $\dot{\mathcal{Q}}_{\Pi,\Pi_{\rm c}} \cap \mathbb{Q}^N$ are in one-to-one correspondence with $\mathcal P_{\Pi,\Pi_{\rm c}}$, where $\mathbb{Q}$ is the set of rational numbers and $\mathcal P_{\Pi, \Pi_{\rm c}}$ is the set of all graph-cover pseudocodewords from all finite graph covers of the 3D-TC factor graph.
\item All vertices of  $\dot{\mathcal{Q}}_{\Pi,\Pi_{\rm c}}$ have rational entries.
\end{enumerate}
\end{proposition}
A similar result was proved in \cite{koe05} for LDPC codes. We give a formal proof for the case of 3D-TCs in Appendix~\ref{sec:prooflem2}. It is inspired by the one in \cite{koe05}, but is a bit more involved because pseudocodewords are defined only indirectly by a linear image of the polytope $\mathcal Q$. Note that our proof does not depend on the detailed set-up of 3D-TCs, so it can be extended to all sorts of turbo-like coding schemes.

When decoding 3D-TCs by solving the linear program in (\ref{eq:LP1}), there is always a vertex of $\dotQPi$ at which the optimum value is attained. We can therefore assume that the LP decoder always returns a vertex and hence, by Proposition~\ref{lem:2}, a (graph-cover) pseudocodeword. Furthermore, the pseudoweight on the AWGN channel of a nonzero pseudocodeword $\boldsymbol{\omega}$ is  defined as \cite{koe05,for01}
\begin{equation} \label{eq:AWGNpseudow}
w^{\rm AWGN}(\boldsymbol{\omega}) = \frac{\lVert \boldsymbol{\omega} \rVert_1^2}{\lVert \boldsymbol{\omega} \rVert_2^2} = 
\frac{\left( \sum_{l=0}^{N-1} \omega_l \right)^2}{\sum_{l=0}^{N-1} \omega_l^2}
\end{equation}
where $\rVert \cdot \rVert_q$ is the $\ell_q$-norm of a vector.


\begin{proposition} \label{lemma_pseudo}
Let $C$ denote a given 3D-TC with interleavers
$\Pi$ and $\Pi_{\rm c}$. For any pseudocodeword $\boldsymbol{\omega}$, the support set $\chi(\boldsymbol{\omega})$ of $\boldsymbol{\omega}$, i.e., the index set of the nonzero
coordinates,  is a stopping set according to \cite[Def.\ 1]{GraRosIsit2010}.
%
Conversely, for any stopping set $\mathcal{S}=\mathcal{S}(\Pi,\Pi_{\rm c})$ of the 3D-TC there exists a pseudocodeword $\boldsymbol{\omega}$ with support set $\chi(\boldsymbol{\omega}) = \mathcal{S}$.
\end{proposition}

\begin{IEEEproof}
This result can be proved in the same manner as the corresponding result for conventional TCs \cite[Lem.\ 2]{ros06}. The proof given in \cite{ros06} is based on the linearity of the subcodes $\bar{C}_{\rm a}$ and $\bar{C}_{\rm b}$ (from the stopping set definition in \cite[Def.\ 1]{ros06}). For 3D-TCs the same proof applies using the linearity of all the three subcodes $\hat{C}_{\rm a}$, $\hat{C}_{\rm b}$, and $\hat{C}_{\rm c}$ from \cite[Def.\ 1]{GraRosIsit2010}. 
\end{IEEEproof}

As a consequence of Proposition~\ref{lemma_pseudo}, it follows that $w^{\rm AWGN}_{\rm min}$  of $C$ is upper-bounded by the $\hmin$ of $C$.

We remark that the $\dmin$ can be computed exactly by solving the \emph{integer} program in (\ref{eq:LP1}) with $\lambda_l=1$ for all $l$, with integer constraints on all the flow variables $\mathbf{f}$ in \eqref{eq:flowSource}--\eqref{eq:flowY}, i.e., $f_l \in \{0,1\}$ for all $l$, and with the constraint $\sum_{l=0}^{N-1} y_l \geq 1$ to avoid obtaining the all-zero codeword (see \cite[Prop.\ 3.6]{Tanatmis+09ValidInequalities}). The exact $\dmin$ of 3D-TCs has not been computed before in the literature, but will be computed later in this paper for several codes. For instance, in \cite{GraRos11}, only estimates of $\dmin$ were provided. Finally, note  that the exact $\hmin$ can be computed in a similar manner using \emph{extended trellis modules} in $T_{\rm x}$ (see \cite{ROS07} for details).


%

\section{Ensemble-Average Pseudoweight Enumerators} \label{sec:average_analysis}

In this section, we describe how to  compute the \emph{ensemble-average pseudoweight enumerator} of  3D-TCs for a given graph cover degree $m$. 


%

Here, we first introduce the concept of a \emph{pseudocodeword vector-weight enumerator (PCVWE)} $\mathcal{P}^{\Cx}_{\mathbf{w},\mathbf{h}}$ of the constituent code $\Cx$. In particular, $\mathcal{P}^{\Cx}_{\mathbf{w},\mathbf{h}}$ is the number of 
\emph{pseudocodewords} in the constituent code $\Cx$,  ${\rm x}={\rm a,b,c}$, of Hamming \emph{vector-weight} $\mathbf{h}=(h_1,\dots,h_m)$ corresponding to input sequences of Hamming \emph{vector-weight} $\mathbf{w}=(w_1,\dots,w_m)$. 
The pseudocodewords of a constituent code $\Cx$ are obtained as follows. Let $\Cx^{(m)}$ denote the degree-$m$ cover of constituent code $\Cx$, which is obtained by concatenating $\Cx$ by itself $m$ times, i.e., 
\begin{displaymath}
\Cx^{(m)} = \bigl\{(\mathbf{x}_{0},\dots,\mathbf{x}_{m-1}): \mathbf{x}_{i} \in \Cx, \forall i \in \{0,\dots,m-1\} \bigr\}.
\end{displaymath}
Now, let
\begin{displaymath}
\mathbf{x}^{(m)}=\bigl(x_{0}^{(0)},\dots,x_{N_{\rm x}-1}^{(0)},\dots,x_{0}^{(m-1)},\dots,x_{N_{\rm x}-1}^{(m-1)}\bigr)
\end{displaymath}
denote a codeword in $\Cx^{(m)}$, where $(x_{0}^{(i)},\dots,x_{N_{\rm x}-1}^{(i)})$ is a codeword in $\Cx$ for all $i$, $0 \leq i \leq m-1$, and $N_{\rm x}$ is the block length of $\Cx$. The corresponding \emph{unnormalized} pseudocodeword is
\begin{equation} \label{eq:pseudocodeword}
\boldsymbol{\omega}(\mathbf{x}^{(m)})= \left( \sum_{i=0}^{m-1} x_{0}^{(i)},\dots,\sum_{i=0}^{m-1} x_{N_{\rm x}-1}^{(i)} \right)
\end{equation}
where addition is integer addition, which means that each component of a pseudocodeword is an integer between $0$ and $m$. The $j$th component $h_j$ of the vector-weight $\mathbf{h} = (h_1,\dots,h_m)$ of the pseudocodeword in (\ref{eq:pseudocodeword}) is the number of components in the pseudocodeword with value $j$, i.e.,
\begin{displaymath}
h_j = \left| \left\{l: \sum_{i=0}^{m-1} x_{l}^{(i)} = j \text{ and } l \in \{0,\dots,N_{\rm x}-1\} \right\} \right|.
\end{displaymath}

The PCVWE of constituent code $\Cx$ can be computed using a nonbinary trellis constructed from the ordinary (information bit-oriented) trellis $T_{\rm x}$. This trellis will be called  the \emph{pseudocodeword trellis} and is denoted by $T_{{\rm x},m}^{\rm PC}$ for constituent code $\Cx$. The procedure to construct $T_{{\rm x},m}^{\rm PC}$ from $T_{\rm x}$ is described below.

\subsection{Constructing $T_{{\rm x},m}^{\rm PC}$ From $T_{\rm x}$}

The pseudocodeword trellis $T_{{\rm x},m}^{\rm PC} = T_{{\rm x},m}^{\rm PC}(V_{{\rm x},m}^{\rm PC},E_{{\rm x},m}^{\rm PC})$, where $V_{{\rm x},m}^{\rm PC}$ is the vertex set and $E_{{\rm x},m}^{\rm PC}$ is the edge set, can be constructed from the trellis $T_{\rm x}$ in the following way. First, 
define the sets
\begin{align*}
\tilde{V}_{{\rm x},m,t}^{\rm PC} &= \overbrace{V_{{\rm x},t} \times V_{{\rm x},t} \times \cdots \times V_{{\rm x},t}}^m \notag \\
\tilde{E}_{{\rm x},m,t}^{\rm PC} &= \bigl\{ ((v_{\rm l}^{(0)},\dots,v_{\rm l}^{(m-1)}),(v_{\rm r}^{(0)},\dots,v_{\rm r}^{(m-1)})): \notag \\
&\;\;\;\;\;\;\;(v_{\rm l}^{(i)},v_{\rm r}^{(i)}) \in E_{{\rm x},t}, \forall i \in \{0,\dots,m-1\} \bigr\} \notag 
\end{align*}
where the time index $t$ runs from $0$ to $I_{\rm x}$ (resp.\ \mbox{$I_{\rm x}-1$}) for the vertices (resp.\ edges). 
The label of an edge $((v_{\rm l}^{(0)},\dots,v_{\rm l}^{(m-1)}),(v_{\rm r}^{(0)},\dots,v_{\rm r}^{(m-1)})) \in \tilde{E}_{{\rm x},m,t}^{\rm PC}$ is the integer sum of the labels of its constituent edges $(v_{\rm l}^{(i)},v_{\rm r}^{(i)}) \in E_{{\rm x},t}$ for all $i$, $0 \leq i \leq m-1$,  which makes the trellis (to be constructed below) nonbinary in general. 

Let $\Psi(\cdot)$ denote a permutation that reorders the components of a vertex $(v^{(0)},\dots,v^{(m-1)}) \in \tilde{V}_{{\rm x},m,t}^{\rm PC} $ according to their labels in a nondecreasing order. As an example, for $m=3$, $\Psi(v_1,v_0,v_2) = (v_0,v_1,v_2)$ and $\Psi(v_2,v_1,v_0) = (v_0,v_1,v_2)$, assuming that vertex $v_i$ has label $i$. Now, define the vertex set $V_{{\rm x},m,t}^{\rm PC}$ by expurgating vertices from the vertex set $\tilde{V}_{{\rm x},m,t}^{\rm PC}$ as follows:
\begin{displaymath}
\begin{split}
V_{{\rm x},m,t}^{\rm PC} &=  \\
&\left\{ \Psi(v^{(0)},\dots,v^{(m-1)}){:}\ (v^{(0)},\dots,v^{(m-1)}) \in \tilde{V}_{{\rm x},m,t}^{\rm PC} \right\}. 
\end{split}
\end{displaymath}

The edge set $E_{{\rm x},m,t}^{\rm PC}$ is defined by expurgating edges from the edge set $\tilde{E}_{{\rm x},m,t}^{\rm PC}$ as follows:
\begin{displaymath}
\begin{split}
E_{{\rm x},m,t}^{\rm PC} = &\bigl\{(\Psi(v_{\rm l}^{(0)},\dots,v_{\rm l}^{(m-1)}),\Psi(v_{\rm r}^{(0)},\dots,v_{\rm r}^{(m-1)})), \\
&\!\forall\, ((v_{\rm l}^{(0)},\dots,v_{\rm l}^{(m-1)}),(v_{\rm r}^{(0)},\dots,v_{\rm r}^{(m-1)})) \in \tilde{E}_{{\rm x},m,t}^{\rm PC} \bigr\}
\end{split}
\end{displaymath}
where all duplicated edges (edges with the same left and right vertex and edge label) are expurgated. The final pseudocodeword trellis is constructed by concatenating the trellis sections $T_{{\rm x},m,t}^{\rm PC} = T_{{\rm x},m,t}^{\rm PC}(V_{{\rm x},m,t}^{\rm PC}, E_{{\rm x},m,t}^{\rm PC})$, $t=0,\dots,I_{\rm x}-1$.

\begin{figure}[!t]
\centering
\includegraphics[height=1.5in]{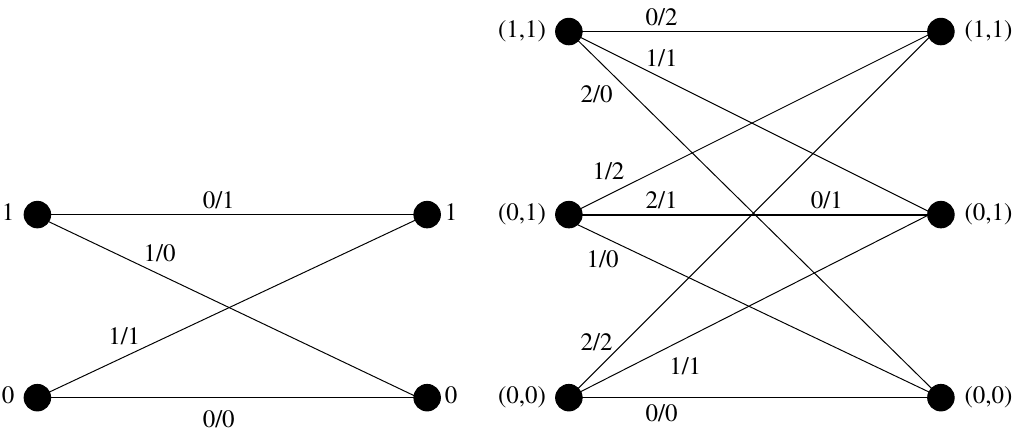}
\caption{The standard trellis section $T_{{\rm x},t}$ (on the left) and the pseudocodeword trellis section $T_{{\rm x},m,t}^{\rm PC}$ (on the right), both being invariant of the time index $t$,  of the accumulator code for $m=2$. Note that for the pseudocodeword trellis section there are two edges with labels $2/1$ and $0/1$, respectively, from the middle vertex to the middle vertex. The vertices of the pseudocodeword trellis section are labeled according to the vertex labeling of the standard trellis section on the left.}
\label{fig:trellis}
\end{figure}

As an example, in Fig.~\ref{fig:trellis}, we show both the standard trellis section $T_{{\rm x},t}$  (on the left) and the pseudocodeword trellis section $T_{{\rm x},m,t}^{\rm PC}$ for $m=2$ (on the right), both being invariant of the time index $t$,  of the accumulator code with generator polynomial $1/(1+D)$. Note that for the pseudocodeword trellis section there are two edges with labels $2/1$ and $0/1$, respectively, from the middle vertex to the middle vertex. 

\begin{lemma} \label{prop:numedges}
For $m=2$, $|V_{{\rm x},m,t}^{\rm PC}| = |V_{{\rm x},t}| + \binom{|V_{{\rm x},t}|}{2}$ and $|E_{{\rm x},m,t}^{\rm PC}| = 2|V_{{\rm x},t}|^2 + |V_{{\rm x},t}|$. 
\end{lemma}

\begin{IEEEproof}
See Appendix~\ref{sec:numedges}.
\end{IEEEproof}

For a $4$-state encoder this means $10$ states, and for an \mbox{$8$-state} encoder this means $36$ states. For an accumulator we only have $3$ states as can be seen in Fig.~\ref{fig:trellis} (the right trellis section). Similar formulas for the number of vertices and edges can be derived for $m > 2$, but are omitted for brevity here.


Now, we define an \emph{equivalence relation} on the set of pseudocodewords as follows. The pseudocodewords $\boldsymbol{\omega}_1$ and $\boldsymbol{\omega}_2$ are said to be \emph{equivalent} if and only if there exists a positive real number $\Delta$ such that $\boldsymbol{\omega}_1 = \Delta \cdot \boldsymbol{\omega}_2$, i.e., they are scaled versions of each other. As a consequence only a single pseudocodeword from an equivalence class can be a vertex of the decoding polytope, which justifies counting equivalence classes only. 
Running a Viterbi-like  algorithm (see Section~\ref{sec:Viterbi} below for details) on the pseudocodeword trellis $T_{{\rm x},m}^{\rm PC}$ constructed above,  will, in general, count pseudocodewords from the same equivalence class. 
However, counting pseudocodewords instead of their equivalence classes does not violate the bounding argument of Section~\ref{sec:finitelength} below, but may lead to a loose bound.  For $m=2$, for instance, these \emph{duplicates} can be removed by a simple procedure which 
removes all terms of $\mathcal{P}^{\Cx}_{\mathbf{w},\mathbf{h}}$ with vector-weights $(\mathbf{w},\mathbf{h}) = ((0,w),(0,h))$. 

As a final remark, the issue of counting pseudocodewords from the same equivalence class is not considered in \cite{abu11,fla09} in the context of LDPC code ensembles.

\subsection{Computing the PCVWE $\mathcal{P}^{\Cx}_{\mathbf{w},\mathbf{h}}$} \label{sec:Viterbi}
The computation of the PCVWE  $\mathcal{P}^{\Cx}_{\mathbf{w},\mathbf{h}}$ for a constituent code $\Cx$ can be performed (for a given cover degree $m$) on the corresponding pseudocodeword trellis $T_{{\rm x},m}^{\rm PC}$, similarly to the computation (on the trellis $T_{\rm x}$) of the input-output weight enumerator. The algorithm performs $I_{\mathrm{x}}$ steps in the trellis. At trellis depth $t$, and for each state $s$, it computes the partial enumerator $\mathcal{P}^{\Cx}_{\mathbf{w},\mathbf{h}}(t,s)$ giving the number of paths in the trellis merging to state $s$ at trellis depth $t$ with input vector-weight $\mathbf{w}$ and output vector-weight $\mathbf{h}$. In particular, $\mathcal{P}^{\Cx}_{\mathbf{w},\mathbf{h}}(t,s)$ is computed from $\mathcal{P}^{\Cx}_{\mathbf{w},\mathbf{h}}(t-1,s^{S}(e))$ by considering all edges $e=(s^{S}(e),s^{E}(e))$ from starting state $s^S(e)$ (at time $t-1$) to ending state $s^E(e)=s$ (at time $t$) according to the dynamic programming principle. Finally, $\mathcal{P}^{\Cx}_{\mathbf{w},\mathbf{h}}= \mathcal{P}_{\mathbf{w},\mathbf{h}}^{\Cx}(I_{\mathrm{x}},s_0)$, where $s_0$ denotes  the all-zero state.

The number of required computations per trellis section is $|E^{\mathrm{PC}}_{\mathrm{x},m,t}| \cdot w_{\max}^m \cdot h_{\max}^m$, where $t=0,\dots,I_{\mathrm{x}}-1$ and $w_{\max}$ (resp.\ $h_{\max}$) is the maximum entry of $\mathbf{w}$ (resp.\ $\mathbf{h}$) that we consider. Note that the computational complexity and the memory requirements scale exponentially with the cover degree $m$ ($m=1$ corresponds to the codeword input-output weight enumerator).


\subsection{Average Pseudoweight Enumerator With Random Puncturing Pattern $\mathbf{p}$}

We assume a random puncturing pattern for $\mathbf{p}$. In particular, 
the puncturing patterns are sampled uniformly at random from the ensemble of puncturing patterns $\mathbf{p}$ with a fraction of $\lambda$ ones. 
Now, using the concept of \textit{uniform interleaver}, the ensemble-average pseudocodeword input-output vector-weight enumerator is
\begin{equation}\label{eq:CWE_3DTC_rand_pseudo}
\begin{split}
\bar{\mathcal{P}}_{\mathbf{w},\mathbf{h}} &=
\sum_{\mathbf{q},\mathbf{q}_\mathrm{a},\mathbf{n}}\frac{\mathcal{P}^{\Ca}_{\mathbf{w},\mathbf{q}_\mathrm{a}}\mathcal{P}^{\Cb}_{\mathbf{w},\mathbf{q}-\mathbf{q}_\mathrm{a}}}{\binom{K}{w_1,w_2,\ldots,w_{m}}} \\ 
&\;\;\;\; \times\frac{\left[ \prod_{i=1}^m \binom{q_i}{n_i} \right] \binom{2K-\sum_{i=1}^m q_i}{2\lambda K-\sum_{i=1}^m n_i}}{\binom{2K}{2\lambda K}} 
\frac{\mathcal{P}^{\Cc}_{\mathbf{n},\mathbf{h}-\mathbf{w}-\mathbf{q}+\mathbf{n}}}{\binom{2\lambda K}{n_1,n_2,\ldots,n_m}} 
\end{split}
\end{equation}
where $\bar{\mathcal{P}}_{\mathbf{w},\mathbf{h}}$ gives the average number (over all interleavers) of unnormalized pseudocodewords of input vector-weight $\mathbf{w}$ and output vector-weight $\mathbf{h}$. In (\ref{eq:CWE_3DTC_rand_pseudo}), 
\begin{displaymath}
{\binom{K}{w_1,w_2,\ldots,w_{m}}} = \frac{K!}{w_1! \cdots w_{m}! (K-\sum_{i=1}^{m} w_i)!},
\end{displaymath}
$\mathbf{q}_{\rm a}$ is the output vector-weight from the constituent code $\Ca$, $\mathbf{q}$ is the total output vector-weight  from the outer turbo code, and $\mathbf{n}$ is input vector-weight for the inner constituent code $\Cc$.

We remark that (\ref{eq:CWE_3DTC_rand_pseudo}) can be seen as a nonbinary version of \cite[Eq.\ (2)]{GraRos11}. 

Now, the ensemble-average pseudoweight enumerator on channel  $\mathcal{H}$ is
\begin{displaymath}
\bar{\mathcal{P}}_{w}^\mathcal{H} = \sum_{\mathbf{w}} \sum_{\substack{ \mathbf{h}: \\ w^{\mathcal{H}}(\mathbf{h}) = w}}  \bar{\mathcal{P}}_{\mathbf{w},\mathbf{h}}
\end{displaymath}
where $w^{\mathcal{H}}(\mathbf{h})$ is the weight metric on  $\mathcal{H}$. For instance, if $\mathcal{H}$ is the AWGN channel, then
\begin{displaymath}
 w^{\mathcal{H}}(\mathbf{h}) = \biggl(\sum_{j=1}^{m} j \cdot h_j \biggr)^2 / \sum_{j=1}^{m} j^2 \cdot h_j 
\end{displaymath}
and if $\mathcal{H}$ is the binary erasure channel, then $w^{\mathcal{H}}(\mathbf{h}) = \sum_{j=1}^{m} h_j$.


\subsection{Average Pseudoweight Enumerator With Regular Puncturing Pattern $\mathbf{p}$}

In a similar fashion as for the case with a random puncturing pattern $\mathbf{p}$, we can modify \cite[Eq.\ (3)]{GraRos11} to arrive at a similar expression (to (\ref{eq:CWE_3DTC_rand_pseudo})) for the ensemble-average pseudocodeword input-output vector-weight  enumerator. Details are omitted for brevity.

\subsection{Finite-Length Minimum Pseudoweight Analysis} \label{sec:finitelength}
The ensemble-average pseudoweight enumerator $\bar{\mathcal{P}}^{\mathcal{H}}_{w}$ can be used to
bound the minimum pseudoweight on $\mathcal{H}$, denoted by $w^{\mathcal{H}}_{\rm min}$,  of the 3D-TC ensemble
in the finite-length regime. In particular, the
probability that a code randomly chosen from the ensemble has
minimum pseudoweight  $w^{\mathcal{H}}_{\rm min} < \wbar$ on $\mathcal{H}$ is upper-bounded by \cite{pfi03}
\begin{equation}\label{eq:probbound}
\mathrm{Pr}(w^{\mathcal{H}}_{\rm min} < \wbar)\leq\sum_{w > 0}^{< \wbar}\bar{\mathcal{P}}^{\mathcal{H}}_w.
\end{equation}
The upper bound in (\ref{eq:probbound}) can be used to obtain a probabilistic lower bound on the minimum pseudoweight of a code ensemble.
For a fixed  value of $\epsilon$, where $\epsilon$ is any positive
value between $0$ and $1$, we define the probabilistic lower bound
with probability $\epsilon$, denoted by $w^{\mathcal{H}}_{{\rm min},{\rm
LB},\epsilon}$, to be the largest real number  $\wbar$ such that the
right-hand side of (\ref{eq:probbound}) is at most $\epsilon$.
This guarantees that ${\rm Pr}(w^{\mathcal{H}}_{\rm min} \geq \wbar ) \geq 1-\epsilon$.

\section{Searching for the Minimum Pseudoweight} \label{sec:searching}

In this section, we present an efficient heuristic to search for low-weight pseudocodewords of 3D-TCs. We use the recently published improved minimum pseudoweight estimation algorithm by Chertkov and Stepanov \cite{che11}. 
In the following, we review that algorithm, restated for 3D-TCs and in a more convenient language.


Recall that the determination of $w_{\min}^{\rm AWGN}$ amounts to minimizing \eqref{eq:AWGNpseudow} over all nonzero vertices of $\dotQPi$. Some important observations allow us to state an equivalent but simpler problem.

Koetter and Vontobel \cite{KoetterVontobel03GraphCovers} already noted that
\begin{IEEEeqnarray}{rCl}
  w^{\rm AWGN}_{\min}
  &=&\!\!
  \min_{\bom \in \dotFPi\setminus \{\boldsymbol{0}\}}\!\! w^{\rm AWGN}(\bom)
  \label{eq:conic}
\end{IEEEeqnarray}
where $\dotFPi$ is the conic hull of $\dotQPi$ (also termed the fundamental cone). The statement follows immediately from the fact that 
$w^{\rm AWGN}(\bom) = w^{\rm AWGN}(\tau\bom)$ for any pseudocodeword $\bom$ and for all $\tau > 0$. The same property allows us to further restrict the search region to the conic section \begin{IEEEeqnarray*}{rCl}
  \Fsec &=& \dotFPi \cap \set{\bom:\:\norm{\bom}_1 = 1}
\end{IEEEeqnarray*}
because every nonzero pseudocodeword may be scaled to satisfy the normalizing condition without changing the pseudoweight. The benefit of this step is twofold: First, in contrast to \eqref{eq:conic} the domain of optimization now is a polytope that can be stated explicitly by means of (in)equalities.
Secondly, minimizing the pseudoweight $w^{\rm AWGN}(\bom)={\norm{\bom}_1^2}/{\norm{\bom}_2^2}$ now is equivalent to maximizing $\lVert\bom\rVert^2_2$, since the numerator is constant on $\Fsec$.

We are thus in the situation of maximizing a convex function ($\norm{\cdot}^2_2$) on a convex polytope. While this is an NP-hard problem in general, the following heuristic proposed in \cite{che11} gives very good results in practice.

For $\bom \in \Fsec$, $\norm{\bom}^2_2 = \norm{\bom-\mathbf1/N}_2^2 +\frac1N$, where $\mathbf1/N=(1,\dotsc,1)/N$, i.e., our goal is to maximize, within $\Fsec$, the distance to the central point $\mathbf1/N$ (the constant $\frac1N$ does not affect maximization). Chertkov and Stepanov proposed to first generate a random point $\bom^{(0)} \neq \mathbf 1/N$ on $\Fsec$, serving as the initial search direction. Then,  the linear program
\begin{displaymath}
  \bom^{(i+1)}
    = \argmax\,(\bom^{(i)} - \mathbf1/N)\bom^T  \text{ subject to }
    \bom \in \Fsec
\end{displaymath}
where $(\cdot)^T$ denotes the transpose of its argument, 
is solved iteratively until the stopping criterion $\bom^{(i+1)}=\bom^{(i)}$ is reached. In each iteration, $\norm{\bom^{(i)}-\mathbf1/N}_1$ increases and therefore $\norm{\bom^{(i)}-\mathbf1/N}_2$ increases as well, and the result is a local maximum. The search is repeated for an arbitrary number of times in different random directions.

In the case of LDPC codes which are covered in \cite{che11}, an explicit description of the polytope in question by means of inequalities is available, thus the fundamental cone can be described explicitly as well by omitting those inequalities which are not tight at $\bom = \boldsymbol{0}$ \cite{KoetterVontobel03GraphCovers}. This is however not the case for the polytope $\dotQPi$ which is only implicitly given as the projection of $\QPi$ onto $\y$. Instead, as we will now show, the cone can be obtained by dropping upper bound constraints on all variables while ensuring that the total flow is equal on all three trellis graphs.

For $\rx = \ra,\rb,\rc$, let $\mathcal Q^\tau_\rx$ be defined as the set of all $\f^\rx \in \mathbb{R}_{\geq 0}^{\abs{\Ex}}$, where $\mathbb{R}$ is the real numbers, satisfying \eqref{eq:flowConservation} and the following modified version of \eqref{eq:flowSource}:
\begin{IEEEeqnarray*}{rCl}
  \sum_{e \in E_{\rx, 0}} f^\rx_e &=& \tau
\end{IEEEeqnarray*}
and let
\begin{equation*}
  \mathcal F =\set{\f=(\f^\ra,\f^\rb,\f^\rc):\: \exists \tau > 0:\:\f^\rx \in \mathcal Q^\tau_\rx \text{ for }\rx = \ra, \rb, \rc}
\end{equation*}
which is, like $\mathcal Q$, the set of all network flows in the trellis graphs, but now with an arbitrary positive total flow $\tau$ instead of $1$. Analogously to $\QPi$, we define $\mathcal F_{\Pi, \Pi_\rc}$ as the set of $(\tilde\y, \f)$ where $\tilde\y \in \mathbb{R}_{\geq 0}^{N+2\lambda K}$ and $\f=(\f^\ra,\f^\rb,\f^\rc) \in \mathcal F$ and additionally \eqref{eq:flowY} is satisfied. The following lemma shows
that the projection of $\mathcal F_{\Pi, \Pi_\rc}$ onto $\y$ indeed yields the fundamental cone of 3D-TC LP decoding.
\begin{lemma} \label{lem:projection}
Let $\dotF$ be the projection of $\mathcal F_{\Pi, \Pi_\rc}$ onto the first $N$ variables. Then, $\dotF = \dotFPi$.
\end{lemma}

\begin{IEEEproof}
See Appendix~\ref{sec:AppendixC}.
\end{IEEEproof}

\section{Numerical Results} \label{sec:numerical}

In this section, we present some numerical results when the interleaver pair $(\Pi,\Pi_{\rm c})$ is taken from the set of all possible interleaver pairs, and when it is taken from the set of pairs of quadratic permutation polynomials (QPPs) over integer rings. Permutation polynomial based interleavers over integer rings for conventional TCs were first proposed 
in \cite{sun05}. These interleavers are  fully algebraic and \textit{maximum contention-free} \cite{tak05}, which makes them very suitable for parallel implementation in the turbo decoder. QPP-based interleavers for conventional TCs were also recently adopted for the 3GPP LTE standard \cite{gpplte}. We remark that for the results below, $\lambda=1/4$ and the regular puncturing pattern $\mathbf{p}=[11000000]$ are assumed. As shown in \cite{GraRos11}, $\lambda=1/4$ gives a suitable trade-off between performance in the waterfall and error floor regions. Finally, we emphasize that all the numerical estimates of the $d_{\rm min}$ and $w^{\rm AWGN}_{\rm min}$ given below are actually also upper bounds on the exact values.

\subsection{Ensemble-Average Results for $K=128$ and $R=1/3$} \label{sec:average}

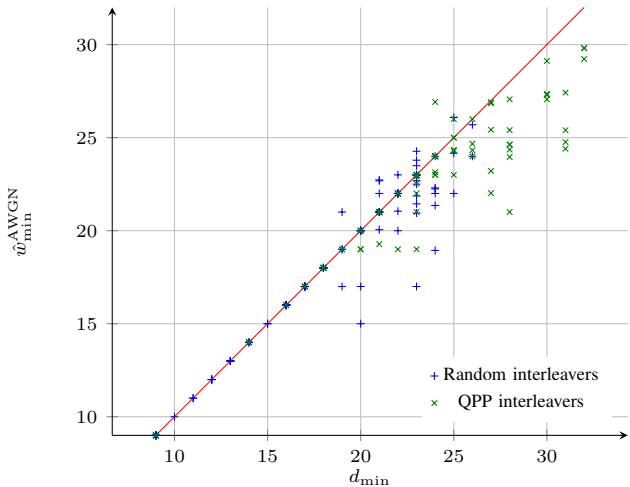
\begin{figure}[tbp]
  \begin{center}
\begin{tikzpicture}
    \begin{axis}[
          font=\scriptsize,
          xlabel=$d_{\min}$,
          ylabel=$\hat w^{\rm AWGN}_{\min}$,
          grid=major,
          axis equal,
          axis x line=bottom,
          axis y line=left,
          legend style={legend pos=south east,draw=none},
x          every axis y label/.append style={yshift=-4mm},
          every axis x label/.append style={yshift=2mm},
          mark options={scale=.75}]
      \addplot[color=red,domain=9:32,forget plot] {x};
      \addplot[only marks,mark=+,color=blue] table[x index=0,y index=1] {Figure5_1.dat};
      \addlegendentry{Random interleavers}

      \addplot[only marks,mark=x,color=green!50!black] table[x index=0,y index=1] {Figure5_2.dat};
      \addlegendentry{QPP interleavers}
    \end{axis}
\end{tikzpicture}
  \end{center}
  \caption{\emph {Estimated $w^{\rm AWGN}_{\rm min}$, using the algorithm from \cite{che08} (which is straightforward to apply to 3D-TCs), and exact $\dmin$ for 3D-TCs with $100$ randomly selected pairs of interleavers (blue plus signs) and with $100$ randomly selected pairs of QPP-based interleavers (green x-marks). The diagonal line gives the trivial upper bound of $\dmin$ on $w^{\rm AWGN}_{\rm min}$ provided by Proposition~\ref{lemma_pseudo}. $K=128$ and $R=1/3$.}}
  \label{fig:K128random}
\end{figure}

In Fig.~\ref{fig:K128random}, we present
the exact $\dmin$ and an estimate of $w^{\rm AWGN}_{\rm min}$ (which is also an upper bound), denoted by $\hat{w}^{\rm AWGN}_{\rm min}$, of unpunctured 3D-TCs with $K=128$ and with $100$ randomly selected pairs of interleavers $(\Pi,\Pi_{\rm c})$ (blue plus signs). The corresponding results with QPP-based interleaver pairs (and with no constraints on the inverse polynomials) are also displayed (green x-marks). For all codes, except $11$, the estimated $w^{\rm AWGN}_{\rm min}$  is at most equal to the $\dmin$. The values of $w^{\rm AWGN}_{\rm min}$  were estimated using the algorithm from \cite{che08} (which is straightforward to apply to 3D-TCs) with a signal-to-noise ratio (SNR) of $2.0$ dB and $500$ evaluations of the algorithm, while the $\dmin$ was computed exactly as described in the second paragraph following the proof of Proposition~\ref{lemma_pseudo}.
 Note that when the $\dmin$ is strictly smaller than the estimated $w^{\rm AWGN}_{\rm min}$ (points above the diagonal line), the estimation algorithm from \cite{che08} was unable to provide an estimate that beats the trivial upper bound provided by 
%
Proposition~\ref{lemma_pseudo}. From the figure, it follows that QPPs give  \emph{better codes} (can provide a higher $\dmin$ and a higher $\hat{w}^{\rm AWGN}_{\rm min}$), and that $w^{\rm AWGN}_{\rm min}$ is strictly lower than $\dmin$ for most codes when the $\dmin$ is large. As a side remark, the algorithm from Section~\ref{sec:searching} gives slightly worse results (the average $\hat{w}^{\rm AWGN}_{\rm min}$ increases by approximately $0.05$) than with the algorithm from  \cite{che08} with the same number of runs ($500$) per instance. However, the algorithm from Section~\ref{sec:searching} is significantly faster.

\subsection{Exhaustive/Random Search Optimizing $w_{\rm min}^{\rm AWGN}$}
In this subsection, we present the results of a computer search for pairs of QPPs with a quadratic inverse for $K=128$, $256$, and $320$ for unpunctured $R=1/3$ 3D-TCs. The objective of the search was to find pairs of QPPs giving a large  estimated $w^{\rm AWGN}_{\rm min}$.  
%
%
To speed up the search, an adaptive threshold on the minimum AWGN pseudoweight $w^{\rm AWGN}_{\rm min}$ was set in the search, in the sense that if a pseudocodeword of AWGN pseudoweight smaller than the threshold was found, then this particular candidate pair of QPPs was rejected.

\begin{figure}[tbp]
  \begin{center}
\begin{tikzpicture}
  \begin{axis}[
        font=\scriptsize,
        height=5cm,
        width=7cm,
        xlabel=Sample number (orderd by increasing $d_{\min}$),
        ylabel={$d_{\min}$, $h_{\min}$, and estimated $w^{\rm AWGN}_{\min}$},
        axis x line=bottom,
        axis y line=left,
        legend style={legend pos=south east,draw=none},
        every axis y label/.append style={yshift=-4mm},
        every axis x label/.append style={yshift=2mm},
        mark options={scale=.75}]
    \addplot+[only marks,mark=o,color=red] table[x expr=\coordindex,y index=0] {Figure6.dat};
    \addplot+[mark=x,color=green!50!black] table[x expr=\coordindex,y index=1] {Figure6.dat};
    \addplot+[only marks,mark=+,color=blue] table[x expr=\coordindex,y index=2] {Figure6.dat};
  \end{axis}
\end{tikzpicture}
  \end{center}
  \caption{\emph {Exact $\dmin$ (red circles), exact $\hmin$ (green x-marks), and $\hat{w}_{\rm min}^{\rm AWGN}$ (blue plus signs) of the $75$ best (in terms of $\hat{w}^{\rm AWGN}_{\rm min}$) QPP-based interleaver pairs for the 3D-TC with input block length $K=128$ and code rate $R=1/3$.}}
  \label{fig:K128}
\end{figure}
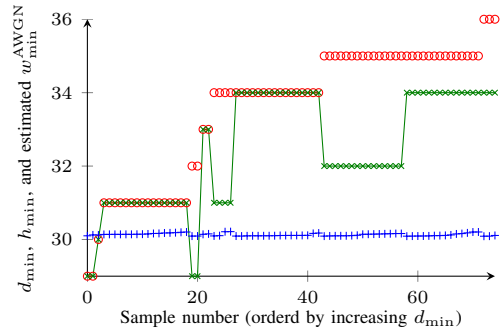

For $K=128$, we performed an exhaustive search over all $2^{17}$ pairs of QPPs (with a quadratic inverse). The minimum AWGN pseudoweight was estimated using the algorithm from \cite{che08} (which is straightforward to apply to 3D-TCs) with an SNR of $1.7$ dB and $500$ evaluations of the algorithm. 
%
%
In Fig.~\ref{fig:K128}, we plot the exact $\dmin$ (red circles), the exact $\hmin$ (green x-marks), and $\hat{w}_{\rm min}^{\rm AWGN}$ (blue plus signs) of the $75$ 3D-TCs with the best $\hat{w}^{\rm AWGN}_{\rm min}$. For each point in the figure, the $x$-coordinate is the sample index (the results are ordered by increasing $\dmin$), while the $y$-coordinate is either the exact $\dmin$, the exact $\hmin$, or $\hat{w}_{\rm min}^{\rm AWGN}$. From the figure, we observe that the best $w^{\rm AWGN}_{\rm min}$ (which is at most $30.2139$) is strictly smaller than the best possible $\dmin$ or $\hmin$. The best possible $\dmin$ was established to be $38$ (exhaustive search), and for this particular code $\hmin=36$, but the estimate of $w^{\rm AWGN}_{\rm min}$ is not among the $75$ best; it is only $29.6042$ (see Table~\ref{table:search} which shows the results of an exhaustive/random search optimizing the $\dmin$ for pairs of QPPs with a quadratic inverse). 
%

For $K=256$, only a partial search has been conducted. The largest found value for $\hat{w}_{\rm min}^{\rm AWGN}$, after taking about $180000$ samples, which is close to $17\%$ of the whole space, is $43.0335$. As for $K=128$, the minimum AWGN pseudoweight was estimated using the algorithm from \cite{che08} (which is straightforward to apply to 3D-TCs) with an SNR of $1.7$ dB and $500$ evaluations of the algorithm. 

For $K=320$, we again performed an exhaustive search over the $2^{18}$ pairs of QPPs with a quadratic inverse. This time we used the algorithm presented in Section~\ref{sec:searching} with $500$ iterations per code.  
%
%
%
%
The largest estimated minimum pseudoweight $\hat w^{\rm AWGN}_{\min}$ that we found was $46.0612$, which is considerably larger than that for the code in Table~\ref{table:search} (which shows the results of an exhaustive/random search optimizing the $\dmin$ for pairs of QPPs with a quadratic inverse).
%
%
%

\begin{table*}[!t]
\begin{minipage}{\linewidth}
\renewcommand\thefootnote{\thempfootnote}
\def\Hline{\noalign{\hrule height 2\arrayrulewidth}}
\par
\begin{center}
\caption{\label{table:search} {Results From an Exhaustive/Random Search for
Pairs of QPPs With $\lambda=1/4$, Both With a Quadratic Inverse, in Which the First QPP
$f(x) = f_1x+f_2x^2 \pmod K$ Generates the TC Interleaver and the Second QPP
$\tilde{f}(x) = \tilde{f}_1x+\tilde{f}_2x^2 \pmod{N_{\rm c}}$ Generates the Permutation in the Patch. Moreover, Terms Like ``SNR'' Are Explained in the Text}} 
\begin{tabular}{lccccclcccc} \Hline
$K$ \T \B & $f_1$ & $f_2$ & $N_\mathrm{c}$ & $\tilde{f}_1$ & $\tilde{f}_2$ &
$\hat{d}_{\rm min}$  & $\hat{w}^{\rm AWGN}_{\rm min}$  & SNR & Evaluations \\
\hline \hline
128\,\footnote{\,Exhaustive search, which implies that the corresponding $\hat{d}_{\rm min}$ is an upper bound on the optimum $\dmin$ (the true optimum $\dmin$ when the estimate $\hat{d}_{\rm min}$ is exact) for this input block length.  \addtocounter{mpfootnote}{+1}\footnotemark\addtocounter{mpfootnote}{-1}\,This is the exact $\dmin$, and we can observe a large gap between $\dmin$ and $w^{\rm AWGN}_{\rm min}$.  \addtocounter{mpfootnote}{+2}\footnotemark\addtocounter{mpfootnote}{-2}\,The QPPs are taken from \cite{GraRos11}.} \T \B & 55  & 96  & 64  & 9  & 16 &  $38$\,\addtocounter{mpfootnote}{+1}\footnotemark[\value{mpfootnote}] 
 &  $29.6042$ & 2.0 dB & 2000 \\ \hline
\addtocounter{mpfootnote}{-1}%
160\,\footnotemark[\value{mpfootnote}]\addtocounter{mpfootnote}{+1} \T \B & 131  & 60  & 80 & 9  & 20  &  $42$\,\footnotemark[\value{mpfootnote}] & $30.0000$ & 1.7 dB & 500\\ \hline
\addtocounter{mpfootnote}{-1}%
192\,\footnotemark[\value{mpfootnote}] \T \B & 35 & 24  & 96  & 11  & 12 &  $46$ & $32.9046$ & 1.7 dB & 500\\ \hline
208\,\footnotemark[\value{mpfootnote}] \T \B & 165 & 182 & 104 & 37 & 26 & $49$ &  $36.3370$ & 1.7 dB & 500\\ \hline
256 \T \B & 239 & 192 & 128 & 37 & 32 & $52$ &  $ 42.7816$ & 1.7 dB & 500 \\ \hline
320 \T \B & 183 & 280 & 160 & 57 & 20 & $58$ & $41.3818$ & 1.7 dB & 500 \\ \hline
\addtocounter{mpfootnote}{1}%
512\,\addtocounter{mpfootnote}{+1}\footnotemark[\value{mpfootnote}]
 \T \B  & 175 & 192 & 256 & 15 & 192 & $67$ & $45.5872$ & 2.0 dB & 500\\ \hline
\end{tabular}
\end{center}
\end{minipage}
\end{table*}






\subsection{Exhaustive/Random Search Optimizing $\dmin$}

We also performed an exhaustive/random search optimizing the $\dmin$ for pairs of QPPs with a quadratic inverse for selected values of $K$ for unpunctured $R=1/3$ 3D-TCs. 
For $K=128$, $160$, $192$, and $208$, the search was exhaustive, in the sense that each pair of interleavers was looked at. In the search, the $\dmin$ was estimated using the triple impulse method \cite{cro04}. The results are given in Table~\ref{table:search} for selected values of $K$, where $f(x)=f_1x+f_2x^2 \pmod K$ generates the TC interleaver, $\tilde{f}(x)=\tilde{f}_1x+\tilde{f}_2x^2 \pmod {N_{\rm c}}$ generates the permutation in the patch, and $\hat{d}_{\rm min}$ and 
$\hat{w}^{\rm AWGN}_{\rm min}$ denote the estimated $\dmin$ and 
the estimated $w^{\rm AWGN}_{\rm min}$, respectively. The estimates of $w^{\rm AWGN}_{\rm min}$ were obtained by using the algorithm from \cite{che08} (which is straightforward to apply to 3D-TCs) with SNR and number of evaluations of the algorithm given in the ninth and tenth column of the table, respectively. 
  Finally, we remark that the codes in the first and second rows, for $K=128$ and $160$, are $\dmin$-optimal, in the sense that there does not exist any pair of QPPs (with a quadratic inverse) giving a $\dmin$ strictly larger than $38$ and $42$, respectively, for the unpunctured 3D-TC. 


\subsection{Ensemble-Average Results for Various $K$ and $R=1/3$}

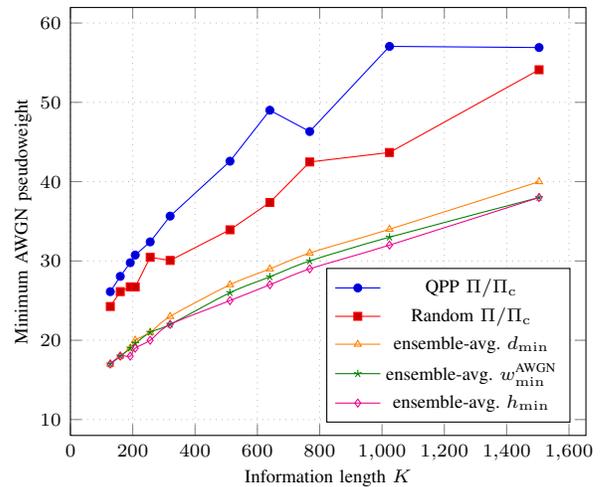
\begin{figure}[htb]
\par
\begin{center}
\pgfplotsset{grid style={help lines,dotted}}
\begin{tikzpicture}
  \begin{axis}[
      font=\scriptsize,
      xlabel=Information length $K$,
      ylabel=Minimum AWGN pseudoweight,
      grid=major,
      legend style={legend pos=south east},
      every axis y label/.append style={yshift=-6mm},
      every axis x label/.append style={yshift=2mm},
      xmin=0,
      ymin=8,
      mark options={scale=.75}
              ]
    \addplot+ table[x=infolength,y=avgpw] {Figure7_2.dat};
    \addlegendentry{QPP $\Pi/\Pi_{\mathrm{c}}$}
    
    \addplot+ table[x=infolength,y=avgpw] {Figure7_1.dat};
    \addlegendentry{Random $\Pi/\Pi_{\mathrm{c}}$}
    
    \addplot+[smooth,orange,mark=triangle] table[x=K,y=dmin] {Figure7_3.dat};
    \addlegendentry{ensemble-avg.~$d_{\min}$}
    
    \addplot+[smooth,draw=green!50!black] table[x=K,y=pseudomin] {Figure7_3.dat};
    \addlegendentry{ensemble-avg.~$w^{\text{AWGN}}_{\min}$}
    
    \addplot+[smooth,draw=magenta!50!magenta,mark=diamond] table[x=K,y=hmin] {Figure7_3.dat};
    \addlegendentry{ensemble-avg.~$h_{\min}$}

  \end{axis}
\end{tikzpicture}
\end{center}

\caption{\label{fig:AverageMinPseudoWeight} The average estimated minimum AWGN pseudoweight for 3D-TCs for different information block lengths $K$, both for QPP-based (with a QPP inverse) and random interleaver pairs. The lower curves show the probabilistic lower bounds on $d_{\min}$, $w_{\min}^{\text{AWGN}}$, and $\hmin$ of the 3D-TC ensemble (for cover degrees of at most $m=2$).}
\end{figure}

In Fig.~\ref{fig:AverageMinPseudoWeight}, we present the average estimated (now using the algorithm from Section~\ref{sec:searching}) minimum AWGN pseudoweight of 3D-TCs for $K=128$, $160$, $192$, $208$, $256$, $320$, $512$, $640$, $768$, $1024$, and $1504$. Both random interleaver pairs and  QPP-based  (with a quadratic inverse) interleaver pairs have been considered. In both cases, we generated $40$ interleaver pairs of each size. For each code we ran $K/10$ trials of the  estimation algorithm described in Section~\ref{sec:searching}. From Fig.~\ref{fig:AverageMinPseudoWeight}, we observe  that the average $\hat{w}_{\rm min}^{\rm AWGN}$ grows with $K$ for both random interleaver pairs and QPP-based interleaver pairs. For all values of $K$, as expected, the average $\hat{w}_{\rm min}^{\rm AWGN}$ is higher for QPP-based interleaver pairs than for random interleaver pairs. 
As a comparison, we have also plotted the corresponding theoretical values $w^{\rm AWGN}_{{\rm min},{\rm LB},0.5}$ from Section~\ref{sec:average_analysis} (using (\ref{eq:probbound})) for graph cover degree $2$. Also, for comparison, we have plotted the corresponding lower bounds on the $\dmin$ and the $\hmin$ using a similar ensemble analysis as the one from  Section~\ref{sec:average_analysis}. For details, we refer the interested reader to \cite{GraRos11,GraRosIsit2010}.  Note that the curves coincide for small values of $K$. The reason that the curve for the probabilistic lower bound on  $w_{\rm min}^{\rm AWGN}$ of the 3D-TC ensemble is higher than the corresponding curve for $h_{\rm min}$ is that the cover degree is limited to $m=2$. In general, the pseudocodewords with support set equal to a small-size stopping set which is not a codeword have a cover degree which is quite large. We would expect that the curve for the probabilistic lower bound on $w_{\rm min}^{\rm AWGN}$ of the 3D-TC ensemble  go further down for $m$ larger than $2$. However, it is currently unfeasible to do the actual computations for larger values of $m$ (both the computational complexity and the memory requirements scale exponentially with $m$). For ease of computation we have used random puncturing patterns $\mathbf{p}$ to compute the curves, while the estimated average values are for regular patterns which in general give better results.

\section{Conclusion} \label{sec:conclu}

In this work, we performed a minimum pseudoweight analysis of pseudocodewords of (relaxed) LP decoding of 3D-TCs, adapting the LP relaxation proposed by Feldman in his thesis for conventional TCs. We proved that the 3D-TC polytope is proper and $C$-symmetric, and made a connection to finite graph covers of the 3D-TC factor graph. This connection was used to show that the support set of any pseudocodeword is a stopping set (as defined in \cite[Def.\ 1]{GraRosIsit2010}), 
and  enabled a finite-length minimum pseudoweight analysis. Furthermore, an explicit description of the fundamental cone of the 3D-TC polytope was given. Finally, both a theoretical and an extensive numerical study of the minimum AWGN pseudoweight of small-to-medium block length 3D-TCs was presented, which showed that 1) typically (i.e., in most cases) when the $d_{\rm min}$ and/or the $h_{\rm min}$ is high,  $w^{\rm AWGN}_{\rm min}$  is strictly smaller than both the $\dmin$ and the $\hmin$ for these codes, and 2) that  $w^{\rm AWGN}_{\rm min}$ grows with the block length, at least for small-to-medium block lengths.
For instance, the exhaustive search for $K=128$ over the entire class of QPP-based interleaver pairs (with a quadratic inverse) revealed  that the best minimum AWGN pseudoweight is strictly smaller than the best minimum/stopping distance.
It is expected that the $w^{\rm AWGN}_{\rm min}$ will dominate the decoding performance for high SNRs.

\appendices
\section{Proof of Proposition~\ref{lem:1}} \label{sec:AppendixA}
We first prove a more general result and then show how it applies to our case.
\begin{lemma}
  \label{lem:csymmetry}
  Let $C_\delta$, $\delta=1,\dotsc, \Delta$, be linear block codes of the same length $N$ and let $C=\cap_{\delta=1}^\Delta C_\delta$. Then, 
  \begin{enumerate}
    \item $\conv(C_\delta)$ is proper and $C_\delta$-symmetric for all $\delta$, and
    \item $\bigcap_{\delta=1}^\Delta \conv(C_\delta)$ is proper and $C$-symmetric.
  \end{enumerate}
\end{lemma}
\begin{IEEEproof}
1) The convex hull $\conv(C_\delta)$ is proper because all codewords are by definition vertices of the polytope. Moreover, because no vertex of the unit hypercube is the convex combination of others, $\conv(C_\delta)$ cannot contain any other integral points. To show $C_\delta$-symmetry, choose $\mathbf{a} \in \conv(C_\delta)$ and $\mathbf{c} \in C_\delta$ arbitrarily. By construction, $\a$ can be written as a convex combination of codewords of $C_\delta$, i.e.,
\begin{equation*}
\a = \sum_{i=1}^{|C_\delta|} \lambda_i \c_i\quad \text{where $\sum_{i=1}^{|C_\delta|} \lambda_i = 1$ and $\lambda_i \geq 0$.}
\end{equation*}
We claim that
\begin{equation}
  \label{eq:symmetry}
  \abs{\a - \c} = \sum_{i=1}^{|C_\delta|} \lambda_i(\c_i \oplus \c) = \sum_{i=1}^{|C_\delta|} \lambda_i \tilde\c_i
\end{equation}
where $\oplus$ denotes  integer
  addition modulo $2$ and $\tilde\c_i$ is the $i$th codeword of $C_\delta$ using a different ordering. This would imply $C_\delta$-symmetry, i.e., if $\a \in \conv(C_\delta)$ and $\c \in C_\delta$, then $\abs{\a-\c} \in \conv(C_\delta)$.

Let $a_j$, $c_j$, and $c_{i,j}$ denote the $j$th coordinate of $\a$, $\c$, and $\c_i$, respectively. The first equality in \eqref{eq:symmetry} follows for $c_j=0$ from
  \begin{IEEEeqnarray*}{rCcCcCcCl}
  \abs{a_j-c_j} &=& a_j = \sum_{i=1}^{|C_\delta|} \lambda_i c_{i,j} &=&
    \sum_{i=1}^{|C_\delta|} \lambda_i (c_{i,j} \oplus c_j )
  \end{IEEEeqnarray*}
and for $c_j=1$ because
  \begin{IEEEeqnarray*}{rCcCcCl}
  \abs{a_j-c_j}
    &=& 1-a_j &=& \sum_{i=1}^{|C_\delta|} \lambda_i (1-c_{i,j} )
    &=& \sum_{i=1}^{|C_\delta|} \lambda_i (c_{i,j} \oplus c_j ).
  \end{IEEEeqnarray*}
The second part of \eqref{eq:symmetry} holds because $\c_i\oplus C_\delta = C_\delta$ due to the linearity of $C_\delta$.

2) Let $\mathcal P=\cap_{\delta=1}^\Delta \conv(C_\delta)$. The properness of $\mathcal P$ for $C$ follows immediately from the properness of $\conv(C_\delta)$, $\delta = 1,\dots,\Delta$, and the definition of $C$. Now, if $\a \in \mathcal P$ and $\c \in C$, then for $\delta=1,\dotsc,\Delta$ we have $\a \in \conv(C_\delta)$ and $\c \in C_\delta$, so by 1) $\abs{\a-\c} \in \conv(C_\delta)$ and thus $\abs{\a-\c} \in \mathcal P$.
\end{IEEEproof}

Now, let $C$ be a 3D-TC. 
By $\tilde C$ we denote the code of length $N+2\lambda K$ obtained by appending the hidden parity bits from $\Ca$ and $\Cb$ which are sent to the patch.
For $\rx = \ra,\rb,\rc$ we define a supercode $\tilde\Cx$ of $\tilde C$ by unconstraining all bits that are not connected to the constituent code $\Cx$, i.e., $\tilde\x \in \tilde\Cx$ if and only if $(\tilde x_{\rho_{\rm x}(0)},\dotsc,\tilde x_{\rho_{\rm x}(N_{\rm x}-1)}) \in \Cx$, where $N_{\rm x}$ is the block length of $\Cx$ and $\rho_{\rm x}(\cdot)$ is defined in \eqref{eq:flowY}, and $\tilde x_i \in \{0,1\}$ for all remaining $i$.
Observe that $\tilde\Ca\cap\tilde\Cb\cap\tilde\Cc = \tilde C$.

Next, define polytopes $\QPi^\rx$ that are obtained from $\QPi$ by dropping in \eqref{eq:flowY} all constraints not corresponding to $\Cx$, and let $\tQPi^\rx$ be the projection of $\QPi^\rx$ onto the $\tilde\y$ variables. Finally, define $\tQPi$ in analogy to $\dotQPi$ as the projection of $\QPi$ onto the first $N+2\lambda K$ variables.

Due to the trellis structure, it is easily seen that $\tQPi^\rx = \conv(\tilde\Cx)$ for $\rx=\ra,\rb,\rc$, and by comparing the polytope definitions we see that $\tQPi^\ra \cap \tQPi^\rb \cap \tQPi^\rc = \tQPi$. Applying Lemma~\ref{lem:csymmetry} shows that $\tQPi$ is both proper and $\tilde C$-symmetric. Now, $C$ is the projection of $\tilde C$ onto the first $N$ variables, and $\dotQPi$ the according projection of $\tQPi$.

To show that $\dotQPi$ is proper, first observe that the projection of any $\tilde{\c} \in \tilde{C}$ onto the first $N$ variables is obviously contained in $\dotQPi$. Conversely, let $\a \in \dotQPi \cap \{0,1\}^N$. By definition there exists $\tilde{\a} \in \tQPi$ such that $\tilde{\a} = (\a, \hat{\a})$. In order to show that $\hat{\a}$ is integral, note that the systematic part of $\tilde{\a}$ is contained in $\a$ and thus integral. Again by the trellis structure, this implies a unique and integral configuration of the flow variables in $T_{\ra}$ and $T_{\rb}$, and consequently also the variables according to the output of those encoders, including the hidden  bits sent to the patch, must be integral. Because $\tQPi$ is proper it follows that $\tilde{\a}\in \tilde{C}$ and thereby also $\a \in C$, which proves properness. Finally, note that $C$-symmetry is trivially preserved by projections, which concludes the proof.

\section{Proof of Proposition~\ref{lem:2}}\label{sec:prooflem2}
Let $\QfPi \subset \mathcal Q$ be the projection of $\QPi$ onto $\f$. We call a flow $\f \in \mathcal Q$ \emph{agreeable} if $\f \in \QfPi$. 
For an agreeable flow $\f$ let $\tilde\y = \tilde\y(\f)$ be the uniquely determined element of $[0,1]^{N+2\lambda K}$ such that $(\tilde{\mathbf y}, \f) \in \QPi$. Analogously, $\y(\f)$ is the projection of $\tilde\y(\f)$ onto the first $N$ variables. Note that $\tilde\y(\f)$ (and $\y(\f)$) can be read off from $\f$ by \eqref{eq:flowY}.

For $\f = (\f^\ra, \f^\rb, \f^\rc) \in \mathcal Q$, but not necessarily $\f \in \QfPi$, we can still use \eqref{eq:flowY} to deduce local values of $\tilde\y$. More precisely, if we define
\begin{equation*}
  \Phi(\rx) = \{ j:\, j = \rho_\rx(\phi_\rx(t,i))\text{ for some }(t,i)\}
\end{equation*}
then for all $j \in \Phi(\rx)$ we can deduce
$\tilde y^\rx_j(\f) = \sum_{\substack{e \in E_{\rx,t}:\\c_i(e)=1}} f^\rx_e$
where $t \in \{0,\dotsc, I_\rx-1\}$ and $i \in \{0,\dotsc,n_{\rx,t}-1\}$ are determined by \eqref{eq:flowY}. This implies that $\f \in \mathcal Q$ is agreeable if and only if 
\begin{equation}
\tilde y^\rx_j(\f) = \tilde y^{\rx'}_j(\f)\label{eq:agreeability}
\end{equation} for all $(j, \rx, \rx')$ such that $j \in \Phi(\rx) \cap \Phi(\rx')$ and $(\rx, \rx') \in \{ (\ra, \rb), (\ra, \rc), (\rb, \rc)\}$, where the first case amounts to the outer interleaver $\Pi$ and the remaining cases are due to the connections to the patch from $\Ca$ and $\Cb$, respectively, via $\Pi_\rc$. We denote the set of these triples $(j,\rx, \rx')$ by $\mathcal A$.

\begin{lemma}
  \label{lem:coverInPolytope}
  The relation $\mathcal P_{\Pi, \Pi_{\rm c}} \subseteq \dotQPi$ holds.
\end{lemma}
\begin{IEEEproof}
Let $\boldsymbol{\omega}(\xm) \in \mathcal P_{\Pi, \Pi_{\rm c}}$ be a graph-cover pseudocodeword of $C$, i.e., there exists a degree-$m$ cover code $C^{(m)}$ of $C$ such that $\xm$ is a codeword of $C^{(m)}$. As before, we can extend $\xm$ to
\[\tilde{\mathbf{x}}^{(m)} = (\tilde x_0^{(0)},\dotsc,\tilde x_{N+2\lambda K-1}^{(0)},\dotsc,\tilde x_0^{(m-1)},\dotsc,\tilde x_{N+2\lambda K-1}^{( m-1)})\]
by appending the parity bits of the copies of $\Ca$ and $\Cb$ that are sent to copies of $\Cc$.

For each $l=0,\dotsc,m-1$, $(\tilde x^{(l)}_0,\dotsc,\tilde x^{(l)}_{N+2\lambda K -1})$ induces via trellis encoding a flow $\f_\bom^{(l)}$ in $\mathcal Q$ with entries only from $\{0,1\}$. In general, $\f_\bom^{(l)}$ is not agreeable because the connections are mixed with different copies 
in the cover graph. However, from the definition of a graph cover we can conclude that
\begin{equation}
  \tilde y_j^{\rm x}(\f_\bom^{(l)}) = \tilde y_j^{\rm x'}(\f_\bom^{(\pi_j(l))})
  \label{eq:agreeabilityCover}
\end{equation}
for all $(j, {\rm x}, {\rm x'}) \in \mathcal A$ and all $l = 0,\dotsc,m-1$, where $\pi_j$ is the corresponding permutation introduced by the graph cover, either (in the case ${\rm x} = {\rm a}$ and ${\rm x'} = {\rm b}$) on connections from an input vertex of $\Gamma(\Ca)$ to a check vertex of $\Gamma(\Cb)$ or (if ${\rm x'} = {\rm c}$) on connections from a parity vertex of $\Ca$ or $\Cb$ to a check vertex of $\Gamma(\Cc)$.

We claim that
\begin{equation}
  \f_\bom = \frac1m \sum_{l=0}^{m-1} \f^{(l)}_\bom
  \label{eq:ffromcover}
\end{equation}
is agreeable and that $\mathbf y(\f_\bom) = \boldsymbol{\omega}(\mathbf x^{(m)})$.

First, note that $\f_\bom$ is a convex combination of elements from the convex set $\mathcal Q$, so $\f_\bom \in \mathcal Q$ as well. To prove agreeability, we verify \eqref{eq:agreeability} for all $(j, {\rm x}, {\rm x'}) \in \mathcal A$:
\begin{IEEEeqnarray*}{rCcCl}
  \tilde y_j^{\rm x}(\f_\bom)
    &=& \sum_{\substack{e\in E_{\rx,t}:\\c_i(e)=1}} f_{\bom,e}^\rx
    &=& \sum_{\substack{e\in E_{\rx,t}:\\c_i(e)=1}}\frac1m \sum_{l=0}^{m-1}
          f^{\rx,(l)}_{\bom, e}
  \\
    &=& \frac1m \sum_{l=0}^{m-1}\tilde y^\rx_j(\f_\bom^{(l)})
    &=& \frac1m \sum_{l=0}^{m-1} \tilde y^{\rm x'}_j(\f_\bom^{(\pi_j(l))})
  \\
    &=& \frac1m \sum_{l=0}^{m-1} \tilde y^{\rm x'}_j(\f_\bom^{(l)})
    &=& \tilde y^{\rm x'}_j(\f_\bom)
\end{IEEEeqnarray*}
where we have used \eqref{eq:flowY}, \eqref{eq:ffromcover}, and \eqref{eq:agreeabilityCover}. The second-to-last equality follows since $\pi_j$ is a permutation of $\{0,\dots,m-1\}$. This shows that $\f_\bom$ is agreeable and thus $\y(\f_\bom)$ is well-defined.

Now, fix $j \in \{0,\dotsc,N-1\}$ and pick any $\rx$ such that $j \in \Phi(\rx)$. Then
\begin{IEEEeqnarray*}{rCcCl}
  y_j(\f_\bom)
    &=& \frac1m \sum_{l=0}^{m-1} \tilde y^\rx_j(\f_\bom^{(l)})
    &=& \frac1m \sum_{l=0}^{m-1} x^{(l)}_j
  \\
    &=& \frac1m \abs{ \{l: \: x_j^{(l)}=1\} }
    &=& \omega_j(\mathbf x^{(m)})
\end{IEEEeqnarray*}
which concludes the proof.
\end{IEEEproof}

Before proving the other direction for rational points, we first show part 2 of Proposition~\ref{lem:2}.

\begin{lemma}
  \label{lem:rationalVertices}
  All vertices of $\dotQPi$ have rational entries.
\end{lemma}
\begin{IEEEproof}
  Let $\y$ be a vertex of $\dotQPi$. Because $\dotQPi$ is a projection of the polytope $\QPi$, and $\QPi$ is the image of $\QfPi$ under a linear map, there exists some vertex $\f$ of $\QfPi$ such that $(\tilde{\mathbf{y}}(\f), \f)$ is also a vertex of $\QPi$ and $\y$ is the projection of $\tilde\y$ onto the first $N$ variables.
  
  Now, $\QfPi$ is a rational polyhedron (i.e., it is defined by (in)equalities with rational entries only), so every vertex of $\QfPi$ is rational \cite[p.\ 123]{NemhauserWolsey88}. Since by \eqref{eq:flowY} each $\tilde{y}_j$ is just a sum of entries of $\f$ for each $j=0,\dotsc,N+2\lambda K-1$, $\tilde\y$ and thus $\y$ must be rational as well. 
\end{IEEEproof}

\begin{lemma}
  \label{lem:rationalFlow}
  For every $\y \in \dotQPi \cap \mathbb Q^N$ there exists a rational point $(\tilde\y, \f) \in \QPi$ such that $\y=\y(\f)$.
\end{lemma}
\begin{IEEEproof}
  Let $\y$ be a rational point of $\dotQPi$. Because $\dotQPi$ is a polytope, $\y$ can be written as a convex combination of vertices of $\dotQPi$, i.e., $ \y=\sum_{k=0}^d \lambda_k \y^k$ where $\lambda_k\ge 0$ for $k=0,\dotsc,d$ and $\sum_{k=0}^d \lambda_k = 1$.
  Furthermore, by Carath{\'{e}}odory's theorem (e.g., \cite[p.\ 94]{Schrijver86LinearIntegerProg}), this is even possible with some $d \leq N$ such that the $\y^k$, $k=0,\dotsc,d$, are affinely independent. Consequently, $\boldsymbol\lambda$ is the unique solution of the system
  \begin{IEEEeqnarray*}{rCl}
    \begin{pmatrix}
      \y^0 & \y^1 & \dotsc & \y^d\\
      1 & 1 & \dotsc & 1
    \end{pmatrix}
    \begin{pmatrix}
      \lambda_0\\
      \vdots\\
      \lambda_d
    \end{pmatrix}
    = \begin{pmatrix} \y \\
1 \end{pmatrix}
  \end{IEEEeqnarray*}
  and by applying Cramer's rule for solving linear equation systems (and Lemma~\ref{lem:rationalVertices} which guarantees that all $\y^k$ have rational entries) we see that $\lambda_k \in \mathbb Q$ for $k=0,\dotsc,d$. Furthermore, the proof of Lemma~\ref{lem:rationalVertices} tells us that for each $\y^k$ there is a rational $\f^k \in \QfPi$ such that $\y^k = \y(\f^k)$. The flow $\f  =\sum_{k=0}^d \lambda_k \f^k$
  satisfies $\y= \y(\f)$ (because $\y(\cdot)$ is linear) and $(\tilde\y(\f), \f) \in \QPi$ is rational, which concludes the proof.
\end{IEEEproof}

Now, we are able to prove the missing counterpart to Lemma~\ref{lem:coverInPolytope}.
\begin{lemma}
  \label{lem:polytopeInCover}
  It holds that $\dotQPi \cap \mathbb{Q}^N \subseteq \mathcal P_{\Pi, \Pi_{\rm c}}$.
\end{lemma}
\begin{IEEEproof}
  Let $\y$ be a rational point of $\dotQPi$. By Lemma~\ref{lem:rationalFlow}, there exist rational $\f \in \QfPi$ and rational $\tilde\y=(\y,\hat\y)$ such that
  $(\tilde\y, \f) \in \QPi$.
  Let $m$ be the least common denominator of the entries of $\f$. Then, $\f_m = m\f$ is a flow with integral values between $0$ and $m$. Applying the flow decomposition theorem \cite[p.\ 80]{Ahuja+93NetworkFlows} in this context guarantees that $\f_m$ can be split up into $m$ binary flows, i.e.,
  \begin{equation}
    \f_m = \sum_{l=0}^{m-1} \f_m^{(l)}
    \label{eq:pathdecomp}
  \end{equation} where $\f_m^{(l)}$ has entries from $\{0,1\}$ and represents a valid path for each trellis $T_\rx$, $\rx= \ra,\rb,\rc$. 
  
  Because $\f \in \QfPi$, we conclude from \eqref{eq:agreeability} that $\tilde y_j^\rx(\f) = \tilde y_j^{\rx'}(\f)$ for all $(j, \rx, \rx') \in \mathcal A$. This is equivalent (by linearity) to $\tilde y_j^\rx(\f_m) = \tilde y_j^{\rx'}(\f_m)$ which by \eqref{eq:pathdecomp} means that $\sum_{l=0}^{m-1}\tilde y_j^\rx(\f_m^{(l)}) =\sum_{l=0}^{m-1}\tilde y_j^{\rx'}(\f_m^{(l)})$. Because all $\f_m^{(l)}$ are $\{0,1\}$-valued, this last equation implies
  \[|\{l:\:\tilde y_j^\rx(\f_m^{(l)})=1\}| = |\{l:\:\tilde y_j^{\rx'}(\f_m^{(l)})=1\}|\]
  and consequently for each $(j,\rx, \rx') \in \mathcal A$ a permutation $\pi_j$ on $\{0,\dotsc,m-1\}$ can be chosen such that $\tilde y_j^\rx(\f_m^{(l)}) = \tilde y_j^{\rx'}(\f_m^{(\pi_j(l))})$ for all $l=0,\dotsc,m-1$. These $\pi_j$ define an $m$-cover $\Gamma^{(m)}(C)$ of $\Gamma(C)$, and by construction 
  \[\x^{(m)} = (x^{(0)}_0,\dotsc,x^{(0)}_{N-1},\dotsc,x^{(m-1)}_0,\dotsc,x^{(m-1)}_{N-1})\]
  is a codeword of $C^{(m)}$, where we define $x^{(l)}_{j} = \tilde y_j^{\rx}(\f^{(l)}_m)$ for the first $\rx$ among $(\ra, \rb, \rc)$ such that $j \in \Phi(\rx)$. Finally, we see that
  \begin{IEEEeqnarray*}{rCl?s}
    \omega_j(\x^{(m)})
      &=& \frac1m \sum_{l=0}^{m-1} x_j^{(l)}
      = \frac1m \sum_{l=0}^{m-1}\tilde y_j^\rx(\f_m^{(l)})
        & (for some x)
    \\
      &=& \frac1m \tilde y_j^\rx(\f_m)
      = \tilde y_j^\rx(\f)
      = y_j
  \end{IEEEeqnarray*}
   (by definition of $\QPi$)
  for any $j=0,\dotsc,N-1$, which shows that $\boldsymbol{\omega}(\x^{(m)}) = \y$.
\end{IEEEproof}

\section{Proof of Lemma~\ref{prop:numedges}}\label{sec:numedges}
The number of vertices in $V^{\rm PC}_{{\rm x},2,t}$ will be the number of distinct ordered $2$-tuples of integers modulo $|V_{{\rm x},t}|$, which is $|V_{{\rm x},t}| + \binom{|V_{{\rm x},t}|}{2}$.

Now, let us consider the number of edges, and in particular, the edges from vertex $\Psi(v_{\rm l},u_{\rm l})$ to vertex $\Psi(v_{\rm r},u_{\rm r})$ in 
$T_{{\rm x},2,t}^{\rm PC}$, where $v_{\rm l}, u_{\rm l}, v_{\rm r}, u_{\rm r} \in V_{{\rm x},t}$.   We have four possible constituent edges to consider, namely the edges
$v_{\rm l}   \xrightarrow{a \ / \ b} v_{\rm r}$, 
$v_{\rm l}   \xrightarrow{c \ / \ d} u_{\rm r}$, 
$u_{\rm l}   \xrightarrow{e \ / \ f} u_{\rm r}$, and 
$u_{\rm l}   \xrightarrow{g \ / \ h} v_{\rm r}$ 
where the labels above the arrows are the input/output labels. Note that $a \neq c$ and $e \neq g$ when $v_{\rm r} \neq u_{\rm r}$. 
Also, the (integer) label of a vertex $v \in V_{{\rm x},t}$ will be denoted by $\ell(v)$. 

\begin{itemize}
\item Case $v_{\rm l} \neq u_{\rm l}$ and $v_{\rm r} \neq u_{\rm r}$: All four constituent edges are distinct, and there will be four edges between the vertices $\Psi(v_{\rm l},u_{\rm l})$ and $\Psi(v_{\rm r},u_{\rm r})$ in $T_{{\rm x},2,t}^{\rm PC}$ with labels $a \! + \! e \ / \ b \! + \! f$, $c \! + \! g \ / \ d \! + \! h$, $g \! + \! c \ / \ h \! + \! d$, and $e \! + \! a \ / \ f \! + \! b$. Since there are only two distinct labels ($a \! + \! e \ / \ b \! + \! f$ and $c \! + \! g \ / \ d \! + \! h$ are always distinct for a minimal trellis/encoder, details omitted for brevity), two of the edges can be removed. 
\item Case $v_{\rm l} \neq u_{\rm l}$ and $v_{\rm r} = u_{\rm r}$: In this case there are only two distinct constituent edges to consider, and there will be two edges between the vertices $\Psi(v_{\rm l},u_{\rm l})$ and $\Psi(v_{\rm r},u_{\rm r})$ in $T_{{\rm x},2,t}^{\rm PC}$  with labels $a \! + \! g \ / \ b \! + \! h$ and $g \! + \! a \ / \ h \! + \! b$ (or $c \! + \! e \ / \ d \! + \! f$ and $e \! + \! c \ / \ f \! + \! d$). 
Since both labels are the same, one of the edges can be removed.

\end{itemize}
In summary, for the first two cases above, we get a total of  $\binom{|V_{{\rm x},t}|}{2} \cdot (2+1+1)$ edges in $T_{{\rm x},2,t}^{\rm PC}$, since there are $\binom{|V_{{\rm x},t}|}{2}$ $2$-tuples $(v_{\rm l},u_{\rm l})$ with $\ell(v_{\rm l}) < \ell(u_{\rm l})$ and $v_{\rm l},u_{\rm l} \in V_{{\rm x},t}$, and two possible values for $v_{\rm r}=u_{\rm r}$ in the second case (the label is either  $a \! + \! g \ / \ b \! + \! h$ or  $c \! + \! e \ / \ d \! + \! f$).
\begin{itemize}
\item Case $v_{\rm l} = u_{\rm l}$ and $v_{\rm r} \neq u_{\rm r}$:
In this case there are again only two distinct constituent edges to consider, and there will be two edges between the vertices $\Psi(v_{\rm l},u_{\rm l})$ and $\Psi(v_{\rm r},u_{\rm r})$ in $T_{{\rm x},2,t}^{\rm PC}$ with labels $a \! + \! c \ / \ b \! + \! d$ and $c \! + \! a \ / \ d \! + \! b$. Since both labels are the same, one of the edges can be removed.
\item Case $v_{\rm l} = u_{\rm l}$ and $v_{\rm r} = u_{\rm r}$:
In this case there is only one distinct constituent edge to consider,  and there will be a single edge between the vertices $\Psi(v_{\rm l},u_{\rm l})$ and $\Psi(v_{\rm r},u_{\rm r})$ in $T_{{\rm x},2,t}^{\rm PC}$ with label $a \! + \! a \ / \ b \! + \! b$ (or $c \! + \! c \ / \ d \! + \! d$).
\end{itemize}
In summary, for the last two cases above, we get a total of  $|V_{{\rm x},t}| \cdot (1+1+1)$ edges in $T_{{\rm x},2,t}^{\rm PC}$, since there are $|V_{{\rm x},t}|$ $2$-tuples $(v_{\rm l},u_{\rm l})$ with $v_{\rm l} = u_{\rm l}$ and $v_{\rm l},u_{\rm l} \in V_{{\rm x},t}$, and two possible values for $v_{\rm r}=u_{\rm r}$ in the fourth case  (the label is either  $a \! + \! a \ / \ b \! + \! b$ or  $c \! + \! c \  /  \  d  \! + \! d$).

In total, there are $4 \binom{|V_{{\rm x},t}|}{2} + 3 |V_{{\rm x},t}| = 2|V_{{\rm x},t}|^2+|V_{{\rm x},t}|$ edges in $T_{{\rm x},2,t}^{\rm PC}$, which is the desired result.

\section{Proof of Lemma~\ref{lem:projection}} \label{sec:AppendixC}
At first we show that $\dotF \subseteq \dotFPi$, so let $\mathbf y \in \dotF$. By definition of $\dotF$, this implies the existence of some $\f=(\f^\ra,\f^\rb,\f^\rc)$,  $\hat\y \in \mathbb{R}_{\geq 0}^{2\lambda K}$, and $\tau > 0$ such that $((\y,\hat\y),\f) \in \mathcal F_{\Pi, \Pi_\rc}$ and $\f^\rx \in \mathcal Q^\tau_\rx$ for $\rx=\ra, \rb, \rc$.
We will show that $(\tilde\y_\tau, \f_\tau) = (\frac1\tau (\y,\hat\y), \frac1\tau\f) \in \QPi$, from which the claim follows because then $\y=\tau\y_\tau$ is a positive multiple of an element of $\dotQPi$.

Conditions \eqref{eq:flowConservation} and \eqref{eq:flowY}, which hold for $((\y, \hat\y), \f)$ by definition of $\mathcal F_{\Pi, \Pi_\rc}$, are invariant to scaling, so they hold for $(\tilde\y_\tau, \f_\tau)$ as well. Because $\f^\rx \in \mathcal Q^\tau_\rx$, it also follows that $\f^\rx_\tau$ satisfies \eqref{eq:flowSource} for all $\rx=\ra,\rb,\rc$.

Equation \eqref{eq:flowSource} also ensures that the total $\f_\tau$-value in the first segment of each trellis $T_\rx$ equals $\frac1\tau \tau = 1$, and because of \eqref{eq:flowConservation} this must hold for all other trellis segments as well. Since $\f_\tau$ is also nonnegative, we can conclude from this that each entry of $\f_\tau$ lies in $[0,1]$. But then also $\tilde\y_\tau \in [0,1]^{N+2\lambda K}$ because each $\tilde{y}_j$, $j=0,\dotsc,N+2\lambda K-1$, is a subset of the total flow through a single segment and thus upper-bounded by $1$, which concludes this part of the proof.

Now, we show that $\dotFPi \subseteq \dotF$. Let $\y \in \dotFPi$. Since $\dotFPi$ is the conic hull of the convex set $\dotQPi$, this implies the existence of some $\tau > 0$ and $\y_{\mathcal Q} \in \dotQPi$ such that $\y = \tau \cdot \y_{\mathcal Q}$.
To $\y_{\mathcal Q}$ then there must exist $\hat\y_{\mathcal Q}$ and $\f_{\mathcal Q}$ such that $((\y_{\mathcal Q},\hat\y_{\mathcal Q}),\f_{\mathcal Q}) \in \QPi$, from which immediately it follows that $((\y = \tau \cdot \y_{\mathcal Q}, \tau \hat\y_{\mathcal Q}), \tau\f_{\mathcal Q}) \in \mathcal{F}_{\Pi, \Pi_\rc}$, and thus $\y \in \dotF$.

\section*{Acknowledgment}

The authors wish to thank the anonymous reviewers for their valuable comments and suggestions that helped improve the presentation of the paper.

\balance



\end{document}